\documentstyle[prd,eqsecnum,aps]{revtex}

\begin{document}

\draft
\title{Resonant particle production with non-minimally
coupled scalar fields\\ in preheating after inflation }
\author{Shinji Tsujikawa$^1$\thanks{electronic
address:shinji@gravity.phys.waseda.ac.jp},
~Kei-ichi Maeda$^{1,2}$\thanks{electronic
address:maeda@gravity.phys.waseda.ac.jp}, and
Takashi Torii$^3$\thanks{electronic
address:torii@th.phys.titech.ac.jp
}\\~}
\address{$^1$ Department of Physics, Waseda University,
Shinjuku, Tokyo 169-8555, Japan\\[.3em]
$^2$ Advanced Research Institute for Science and Engineering,\\
Waseda University, Shinjuku, Tokyo 169-8555, Japan\\[.3em]
$^3$ Department of Physics, Tokyo
Institute of Technology, Meguro, Tokyo 152, Japan\\~}
\date{\today}
\maketitle
\begin{abstract}
We investigate a resonant particle production of a
scalar field $\chi$ coupled non-minimally to a spacetime
curvature $R$ ($\xi R \chi^2$) as well as to an inflaton
field $\phi$ ($g^2\phi^2\chi^2$).
In the case of
$g~\mbox{\raisebox{-1.ex}{$\stackrel
     {\textstyle<}{\textstyle \sim}$}}~3 \times 10^{-4}$, $\xi$ effect
assists $g$-resonance in certain parameter regimes.
However,
for $g~\mbox{\raisebox{-1.ex}{$\stackrel
     {\textstyle>}{\textstyle\sim}$}}~3
\times 10^{-4}$, $g$-resonance is not
enhanced by
$\xi$ effect because of $\xi$ suppression effect as well as
a back reaction effect.
If $\xi \approx -4$, the maximal fluctuation of produced
$\chi$-particle is $\sqrt{\langle \chi^2
\rangle}_{max} \approx 2 \times 10^{17}$ GeV for $g
~\mbox{\raisebox{-1.ex}{$\stackrel{\textstyle<}{\textstyle \sim}$}}~1 \times 10^{-5}$, which is larger than the minimally coupled case with $g \approx 1 \times 10^{-3}$.

\end{abstract}
\pacs{98.80.Cq, 05.70.Fh, 11.15.Kc}

\baselineskip = 24pt

%%%%%%%%%%%%%%%%%%%%%%%%%%%%%%%%%%%%%%%%%%%%%%%%%%
%
%                                                 %
\section{Introduction}                            %
%                                                 %
%%%%%%%%%%%%%%%%%%%%%%%%%%%%%%%%%%%%%%%%%%%%%%%%%%
\ \ In an inflationary scenario, most of the elementary
particles in the Universe were created during the stage of
reheating after inflation \cite{Kolb}.
During the inflationary stage, an inflaton field slowly
rolls down to a  minimum of its potential.
A reheating process turns on when the inflaton field begins
to oscillate around
the minimum of its potential.
The original version of reheating scenario was first
considered in the context of a new inflationary model in
\cite{Dolgov,Abbott}.
The particles are created through an interaction term
between the inflaton and some fields. A phenomenological
decay term to describe the fact that  the inflaton field
decays to other lighter particles (radiation)
 is included in   the equation of the inflaton  field $\phi$,
 and the energy
of the inflaton is transferred to their thermal energy.
According to this scenario, reheating temperature is
determined by the decay rate
$\Gamma$ but not the initial value of
$\phi$. The value of $\Gamma$ is
constrained to be small by the perturbation  theory such as
$\Gamma<10^{-20} M_{\rm PL}$, where
$M_{\rm PL}$ is the Planck mass, so the reheating
temperature is estimated as $T_r <10^9$ GeV. This
temperature is not sufficient in order to produce baryon
asymmetry based on Grand Unified Theories (GUTs).

Recently, however, it has been recognized that
reheating process begins by a parametric amplification
of scalar particles\cite{TB,KLS1,Boy1,Yoshi}.
This initial evolutionary phase, which is called  {\it
preheating} stage,  provides an explosive particle
production and must be discussed separately from the
perturbative decay of inflaton. There are many works about
the preheating stage based on analytical
investigations\cite{KLS2,Green,Kaiser,Son} as well as on
numerical studies\cite{KT1,KT2,PR}.  The important feature with
the existence of preheating stage is that the maximal value of
produced fluctuation can be so large that it would result in
a non-thermal phase transition
\cite{KLRT} and make baryogenesis at the GUT
scale possible\cite{KLR},  although the baryogenesis might be
important in much lower energy scale, i.e. the electro-weak
scale\cite{EWbaryon}.

So far, we know two possible preheating scenarios: one is that
the inflaton field $\phi$ itself is transformed into many
$\phi$-particles from a coherent inflationary phase through a
self-interaction such as
$\lambda
\phi^4$, and the other is that another field, e.g. a scalar
field $\chi$ is created through some coupling with the inflaton
such as $g^2\phi^2\chi^2$.
Both cases were first discussed in \cite{KLS1} using Hartree
mean-field approximation.
For the former case, the preheating was studied\cite{Boy1,Boy2}
by making use of closed time path formalism
\cite{Jor,Cal,Baa,CTP}, since the preheating is
an essentially  non-equilibrium state.
 They analyzed the non-perturbative evolution of
the inflaton fluctuations for the self-interacting  massive
inflaton by the method of the $O(N)$-vector model in the large
$N$ limit as well as by the Hartree factorization model, both of
which are mean-field approximations.  The $O(N)$-vector
model has an advantage to deal with the continuous symmetry
while the Hartree factorization model is suitable to treat
the discrete symmetry.
To confirm the mean-field approximation, the fully
non-linear numerical simulation including the scattering
effect of created particles was also performed in
\cite{KT1} for the simple
$\lambda\phi^4$ model, finding that  the variance
$\langle\delta\phi^2\rangle
\sim 10^{-7} M_{\rm PL}^2$ can be produced by the
non-perturbative process of inflaton decay.
They treated the scalar field fluctuations as classical ones,
which would be justified because the fluctuations with rather
low momenta are mainly produced by the parametric resonance.
They found
almost the same result as that by the mean-field approximation.

As for the case only with a massive-inflaton potential 
(${1 \over
2}m^2\phi^2)$,   the production of inflaton fluctuation will
not be expected by a parametric resonance. Hence one usually
introduces another scalar field $\chi$
 coupled to the inflaton field $\phi$ through an
 interaction such
as $g^2\phi^2\chi^2$. For the production of $\phi$-particles in
self-interacting  case,  the coupling is constrained to be
$\lambda \sim 10^{-12}$ from the observation of COBE, while
the coupling constant $g$ in the present case is a free
parameter.  Hence we may find a  higher production of
$\chi$-particles depending on the value of $g$. As for
analytical investigation of $\chi$-field evolution, Kofman,
Linde and Starobinsky\cite{KLS1,KLS2} developed  the
consistent theory of preheating based on Mathieu equation for
$\chi$-field. There are several numerical works  devoted
to the  evolution of $\chi$-field by fully non-linear
simulations\cite{KT2,PR}.  A parametric resonance turns on
from the broad resonance regime in certain values of the
coupling $g$. The structure of resonance will be modified a
lot in the expanding Universe compared with that in Minkowski
space. The amplitude of the coherent $\phi$ field decreases
adiabatically because of the expansion of the Universe, and
eventually the resonance terminates.  In the case that the
coupling $g$ is small and the resonance  band is narrow from the
first stage of preheating, an efficient resonance will not be
expected. For the large coupling constant, the resonance band is
broad in the beginning, and then  we find a considerable
production of $\chi$-particles. 
In this case, $\chi$-field crosses many instability and stability bands
even within one oscillation of the inflaton field, and stochastic resonance occurs.
A significant amplification occurs in the broad resonance regime,
but in certain values of coupling constant $g$, production via
such a resonance will be suppressed by a back reaction of created
particles.  When such a back reaction is taken into
account, a coherent oscillation of the inflaton field
is broken due to the increase of the effective mass of inflaton,
and the energy of $\chi$-field can not exceed much larger than that of $\phi$-field at the final stage of preheating.
Moreover, rescattering effect, which will also restrict
$\chi$-particles production, becomes important at the final
stage of preheating.

Recently, another interesting mechanism for preheating
called a geometric reheating  model has been  proposed by
Bassett and Liberati
\cite{BL}. They investigated the case that $\chi$ field is
non-minimally coupled to spacetime curvature $R$ for massive
inflaton model and found that an explosive
amplification of $\chi$ will be possible if the coupling
constant
$\xi$ is negative\cite{comment1}.
This is just because of unstable modes with a negative
coupling\cite{instability}. In this sense, it is different
from ordinary parametric resonance. In particular, they
studied how the homogeneous $\chi$ field is amplified
and found that GUT scale  gauge boson with mass
$m_{G} \sim 10^{16}$ GeV
can be produced in preheating stage when $\xi$ is 
large negative.

A natural question may arise: If we combine both  effects of 
$g$-resonance and
$\xi$-resonance, whether we find  more effective production of
$\chi$-particles, since small coupling $g$ does not provide a
sufficient production. It is of interest  how the
$g$-resonance picture is modified by taking into account the
effect of geometric reheating.
If the assisting mechanism works and final abundance of
$\chi$ becomes rather larger, it may change the ordinary
preheating picture and affect  the non-thermal phase
transition and baryogenesis.  We will discuss such a
combined resonance in detail.
We also study the evolution of $\chi$ field fluctuation
summed by all possible momentum modes
in detail including the back reaction effect, because  the
previous study is mainly devoted to the growth of zero momentum
mode at the initial stage of preheating\cite{BL}.

This paper is organized as follows.
In the next section, we introduce basic equations in the
non-minimal  preheating that includes ordinary $g$-resonance.
The results of $g=0, \xi>0$ case is presented in Sec. III.
We study the analytical estimation as well as numerical
results, and compare them with each other.
In Sec. IV, we analyze $g=0, \xi<0$ case.
The structure of negative coupling instability is
investigated. In Sec. V, the combined effect of $g$ and
$\xi$ resonance is studied. We present that at which values
of $g$ and $\xi$ parametric resonance becomes most efficient.
Finally, we give our discussions and conclusions in the
final section.

%%%%%%%%%%%%%%%%%%%%%%%%%%%%%%%%%%%%%%%%%%%%%%%%%%
%%%%%%%%%%
%                                                          %
\section{Basic equations}    %
%                                                          %
%%%%%%%%%%%%%%%%%%%%%%%%%%%%%%%%%%%%%%%%%%%%%%%%%%
%%%%%%%%%%

%%%%%%%%%%%%%
We consider a model that an inflaton field $\phi$ is
interacting with a scalar field $\chi$, which is
non-minimally coupled with the spacetime curvature $R$,
\begin{eqnarray}
{\cal L} = \sqrt{-g} \left[ \frac{1}{2\kappa^2}R
   -\frac{1}{2}(\nabla \phi)^2
   -V(\phi)
   -\frac{1}{2}(\nabla \chi)^2
   -\frac{1}{2} m_{\chi}^2 \chi^2
   -\frac{1}{2} g^2 \phi^2 \chi^2
   - \frac{1}{2}\xi R \chi^2 \right],
\label{B1}
\end{eqnarray}
%%%%%%%%%%%%%
where $\kappa^{2}/8\pi \equiv G =M_{\rm PL}^{-2} $ is
Newton's gravitational constant, $g$ and $\xi$ are coupling
constants, and
$m_{\chi}$ is a mass of the $\chi$ field.
$V(\phi)$ is a potential of the inflaton field. 
In this paper we adopt the quadratic potential
%%%%%%%%%%%%%%
\begin{eqnarray}
V(\phi)=\frac{1}{2}m^2\phi^2,
\end{eqnarray}
%%%%%%%%%%%%%%
where $m$ is a mass of the inflaton field, and we use the
value of $m=10^{-6} M_{\rm PL}$ that is obtained by fitting
density perturbations to COBE data.

Since non-minimal coupling between the spacetime curvature
$R$ and $\chi$ field makes the basic
equations complicated,
it is convenient to transform  action (\ref{B1})
into  the Einstein frame by a conformal transformation
\cite{FM}.
We make the conformal transformation as follows:
%%%%%%%%%%%%%%
\begin{eqnarray}
\hat{g}_{\mu \nu}=\Omega^2 g_{\mu \nu} ,
\label{B2}
\end{eqnarray}
%%%%%%%%%%%%%%
where $\Omega^2  \equiv 1-\xi\kappa^2 \chi^2$.
The Lagrangian density in the Einstein frame becomes
%%%%%%%%%%%%%%%
%%%%%%%%%%%%%%%
\begin{eqnarray}
{\cal L} = \sqrt{-\hat{g}} \left[ \frac{1}{2\kappa^2} \hat{R}
   -\frac{1}{2\Omega^2}(\hat{\nabla} \phi)^2
   -\frac{1}{2\Omega^4}m^2\phi^2
   -\frac{1}{2} F^2 (\hat{\nabla} \chi)^2
   -\frac{1}{2\Omega^4}(m_{\chi}^2+g^2\phi^2)\chi^2 \right],
\label{B3}
\end{eqnarray}
%%%%%%%%%%%%%%%
%%%%%%%%%%%%%%%
where variables with a caret  denote those in the
Einstein frame, and
%%%%%%%%%%%%%%%
\begin{eqnarray}
F^2 = \frac{1-(1-6\xi) \xi\kappa^2 \chi^2 }
{(1-\xi\kappa^2 \chi^2)^2} .
\label{B4}
\end{eqnarray}
%%%%%%%%%%%%%%%
To make the kinetic term of the $\chi$ field canonical,
we define a new scalar field $X$ as
%%%%%%%%%%%%%%%
\begin{eqnarray}
X \equiv \int F(\chi) d\chi ,
\label{B5}
\end{eqnarray}
%%%%%%%%%%%%%%%
by which the Lagrangian density is now
%%%%%%%%%%%%%%%
%%%%%%%%%%%%%%%
\begin{eqnarray}
{\cal L} = \sqrt{-\hat{g}} \left[ \frac{1}{2\kappa^2} \hat{R}
   -\frac{1}{2\Omega^2}(\hat{\nabla} \phi)^2
   -\frac{1}{2\Omega^4}m^2\phi^2
   -\frac{1}{2} (\hat{\nabla} X)^2
   -\frac{1}{2\Omega^4}(m_{\chi}^2+g^2\phi^2)\chi^2(X) \right].
\label{B6}
\end{eqnarray}
%%%%%%%%%%%%%%%

Since we are interested 
in a preheating after inflation, as usual, we
shall assume that the spacetime and the inflaton
$\phi$ give a classical background and the scalar field $\chi$
is treated as a quantum field on that background.
In the present model, however, there is some problem.
We have performed a conformal transformation, which conformal
factor $\Omega^2$ includes quantum variable $\chi^2$.
Then, in order not to discuss quantum gravity, the conformal factor
should be replaced with some expectation value.
Here, we regard the conformal factor $\Omega^2$ as
$1-\eta$\cite{comment2}, where 
%%%%%%%%%%%%%%%
\begin{eqnarray}
\eta \equiv \xi \kappa^2 \langle\chi^2\rangle.
\label{B11}
\end{eqnarray}
%%%%%%%%%%%%%%%
$\langle \chi^2 \rangle$ corresponds to  
the number density of the created
$\chi$-particle, where $\langle \cdots \rangle$ denotes an
expectation value of some functional of  $\chi$ (or $X$).

With such a transformation, we assume that the spacetime and
the inflaton field $\phi$ are spatially homogeneous, 
and adopt the flat Friedmann-Robertson-Walker metric 
as the background spacetime;
%%%%%%%%%%%%%%%
\begin{eqnarray}
d\hat{s}^2 = -dt^2 + \hat{a}^2(t) d {\bf x}^2.
\end{eqnarray}
%%%%%%%%%%%%%%%
Hereafter, as we argue only in the Einstein frame, we drop a
caret. The evolution of the scale factor $a$ yields
%%%%%%%%%%%%%%%
\begin{eqnarray}
\left(\frac{\dot{a}}{a}\right)^2 =
   \frac{\kappa^2}{3}
     \left[ \frac{1}{2(1-\eta)} \dot{\phi}^2
   +\frac{1}{2(1-\eta)^2} m^2\phi^2
   +\frac{1}{2}\langle (\nabla X)^2 \rangle
   +\frac{1}{2 (1-\eta)^2 }
      (m_{\chi}^2+g^2\phi^2)\langle\chi^2\rangle
\right],
\label{B9}
\end{eqnarray}
%%%%%%%%%%%%%%%
where a dot denotes a derivative with respect to time
coordinate $t$.
%%%%%%%%%%%%%%%

The evolution of the inflaton field $\phi$
is described by
%%%%%%%%%%%%%%%
\begin{eqnarray}
\ddot{\phi}+\left( 3\frac{\dot{a}}{a}+ \frac{\dot{\eta}}{1-\eta}
\right) \dot{\phi}
+\frac{1}{1-\eta}
 (m^2+g^2\langle\chi^2\rangle) \phi=0.
\label{B10}
\end{eqnarray}
%%%%%%%%%%%%%%%
Note that the fluctuation of the inflaton field $\delta \phi$
is not considered
since we assume that the inflaton field is spatially
homogeneous. However, the growth of the $\phi$ fluctuation
would be expected to appear as $\chi$ field is amplified. Our
investigations are limited in the sense that rescattering
between $\chi$ and $\delta\phi$ fluctuation is not included.

In Eq.~$(\ref{B10})$, if we introduce a new scalar field
$\varphi$ such as
%%%%%%%%%%%%%%%
\begin{eqnarray}
\varphi \equiv b^{3/2} \phi,
\label{B12}
\end{eqnarray}
where $b^3 =a^3/(1-\eta)$,
then the field $\varphi$ obeys the following equation:
%%%%%%%%%%%%%%%
\begin{eqnarray}
\ddot{\varphi}
   +\left[ \frac{1}{1-\eta}
     \left(m^2+g^2\langle\chi^2\rangle \right)
   -\frac34\left(\frac{2\ddot{b}}{b}
+\frac{\dot{b}^2}{b^2}\right)
   \right] \varphi =0 .
\label{B13}
\end{eqnarray}
%%%%%%%%%%%%%%%
Note that in the case of minimal coupling $\xi=0$,
$\eta$ vanishes, hence
the coherent oscillation
of $\varphi$ is broken only by $g^2\langle\chi^2\rangle$
term.
As $\langle\chi^2\rangle$ grows, the effective  mass of the
inflaton
$m^2_{eff} = m^2+g^2\langle\chi^2\rangle$
gets large, i.e., oscillation becomes rapid.
This effect, which is called back reaction effect
to the inflaton field, suppresses the resonant particle
creation. On the contrary, when $\xi$ effect is taken into
account,
$1-\eta$ is decreasing as $\langle\chi^2\rangle$ grows,
hence one may expect this effect also changes the
coherent oscillation of $\varphi$.
However, as we will see later, this is not the case and we can
neglect $\xi$ effect on the inflaton field in most cases.

Next, let us consider the equation of $X$ field.
The Heisenberg equation of motion is derived from (\ref{B6}):
%%%%%%%%%%%%%%%
\begin{eqnarray}
\ddot{X}
   +3\frac{\dot{a}}{a}\dot{X}-\partial_i\partial^i X
   -\frac{\partial}{\partial X} \left[ \frac{\dot{\phi}^2}
    {2 \Omega^2}
   -\frac{1}{2\Omega^4}m^2 \phi^2
   -\frac{1}{2\Omega^4}
      (m_{\chi}^2+g^2\phi^2) \chi^2
    \right]=0,
\label{B14}
\end{eqnarray}
%%%%%%%%%%%%%%%
where an index with a roman character denotes space coordinates.
In order to study a quantum particle creation of $\chi$ fields,
we make the following
mean field approximation with respect to $X$, which provides us a
linearized equation for a quantum field $X$ :
%%%%%%%%%%%%%%%
\begin{eqnarray}
\ddot{X}
   +3\frac{\dot{a}}{a}\dot{X}-\partial_i\partial^i X
   -
    \nabla_X^2\left[ \frac{\dot{\phi}^2}
    {2  \Omega^2  }
   -\frac{1}{2 \Omega^4}m^2 \phi^2
   -\frac{1}{2  \Omega^4}
      (m_{\chi}^2+g^2\phi^2)  \langle\chi^2 \rangle
    \right]  X =0,
\label{B15}
\end{eqnarray}
where $\nabla_X=2\sqrt{\langle X^2\rangle} ~\partial / \partial
\langle X^2\rangle$.  {}From the
relation (\ref{B5}), we have also assumed
\begin{equation}
d \left( \sqrt{\langle X^2\rangle} \right) =
\sqrt{\langle F^2\rangle} ~d \left(\sqrt{\langle \chi^2\rangle}
\right) ,
\label{B8}
\end{equation}
where $\langle F^2\rangle = [1-(1-6 \xi) \eta ]/(1-\eta)^2$.
%%%%%%%%%%%%%%%
Expanding the  scalar fields $X$   as
%%%%%%%%%%%%%%%
\begin{eqnarray}
X=\frac{1}{(2\pi)^{3/2}} \int \left(a_k X_k(t)
 e^{-i {\bf k} \cdot {\bf x}}+a_k^{\dagger} X_k^{*}(t)
 e^{i {\bf k} \cdot {\bf x}} \right) d^3{\bf k},
\label{B21}
\end{eqnarray}
%%%%%%%%%%%%%%%
where $a_k$ and $a_k^{\dagger}$ are the annihilation and
creation operators, we find that
  $X_k$ obeys the following equation of motion:
%%%%%%%%%%%%%%%
\begin{eqnarray}
\ddot{X}_k
   +3\frac{\dot{a}}{a}\dot{X}_k
   +\left[\frac{k^2}{a^2} +
      G(\langle\chi^2\rangle) \right] X_k =0,
\end{eqnarray}
\label{B17}
%%%%%%%%%%%%%%%
with
\begin{eqnarray}
G(\langle\chi^2\rangle) &\equiv &
   \frac{1}{(1-\eta)\left[1-(1-6\xi)\eta\right]}
   \biggl[ (1+3\eta)(m_{\chi}^2 +g^2\phi^2)
      +2\xi\kappa^2m^2\phi^2
      -(1-3\eta)\xi\kappa^2 \dot{\phi}^2  \nonumber \\
   & &
      +\frac{\eta}{1-\eta}
      \frac{4+(1-5\eta)(1-6\xi)}{1-(1-6\xi)\eta}
      \left\{(1+\eta)(m_{\chi}^2+g^2\phi^2)
         +2\xi\kappa^2m^2\phi^2
-(1-\eta)\xi\kappa^2\dot{\phi}^2
       \right\} \biggr].
\label{B18}
\end{eqnarray}
%%%%%%%%%%%%%%%
%%%%%%%%%%%%%%%
The expectation values of $X^2$ and $\chi^2$ are expressed as
%%%%%%%%%%%%%%%
\begin{eqnarray}
\langle X^2 \rangle= \frac1{2\pi^2} \int k^2|X_k|^2 dk,
\;\;\;\;
\langle \chi^2 \rangle = \frac1{2\pi^2} \int k^2|\chi_k|^2 dk.
\label{B7}
\end{eqnarray}
%%%%%%%%%%%%%%%
Introducing the function $Y_k =a^{3/2} X_k$, instead of $X_k$,
we find
%%%%%%%%%%%%%%%
\begin{eqnarray}
\ddot{Y_k} +\omega_k^2 Y_k =0,
\label{B19}
\end{eqnarray}
%%%%%%%%%%%%%%%
where
%%%%%%%%%%%%%%%
\begin{eqnarray}
\omega_k^2 =\frac{k^2}{a^2} +  G(\langle\chi^2\rangle)
                     -\frac34\left(\frac{2\ddot{a}}{a}+
                     \frac{\dot{a}^2}{a^2}\right),
\label{B20}
\end{eqnarray}
%%%%%%%%%%%%%%%
which is a time dependent frequency of $Y_k$.

At the first stage of preheating when $\chi$-particles are
produced  by quantum
 fluctuation, we find the occupation number in $Y_k$-state
by the Bogoliubov transformation as
%%%%%%%%%%%%%%%
\begin{eqnarray}
n_k = \frac{\omega_k}{2} \left( |Y_k|^2 +\frac{|\dot{Y}_k|^2}
{\omega_k^2} \right) - \frac{1}{2}.
\label{B23}
\end{eqnarray}
%%%%%%%%%%%%%%%
However, after many $\chi$-particles are created and each mode
is amplified, the $\chi_k$ field could be regarded as the
classical field in a good approximation\cite{KT2,PR}.
We may not have to use Eq.~(\ref{B23}) at the classical stage.
Khlebnikov and Tkachev developed the semiclassical
description of fluctuation produced by the inflaton decay
\cite{KT1}. In fact, they studied the inflaton decay by the
classical equation of motion and performed a fully nonlinear
calculation. To give  the initial conditions, we
follow their approach, i.e. the initial distribution for
$Y_k$ is described as
%%%%%%%%%%%%%%%
\begin{eqnarray}
P[Y_k ; t=0] = \sqrt{\frac{2\omega_k(0)}{\pi}} \exp\left[
-2\omega_k(0) |Y_k|^2 \right],
\label{B24}
\end{eqnarray}
%%%%%%%%%%%%%%%
and $\dot{Y}_k$ is correlated to $Y_k$ as
\begin{eqnarray}
\dot{Y}_k =-i \omega_k(0) Y_k .
\label{B25}
\end{eqnarray}
%%%%%%%%%%%%%%%
We investigate the
$\langle \chi^2 \rangle$ evolution with those initial
conditions as the semiclassical problem.

{}From Eq.~(\ref{B18}), we  easily find that
properties of the preheating in the non-minimal coupling theory
are quite different from ordinary $g$-resonance. First,
$\xi\kappa^2m^2\phi^2$ term and $\xi\kappa^2 \dot{\phi}^2$
term cause the resonance
as well as interaction term $g^2\phi^2$ with inflaton field.
These different types of resonant terms either strengthen or
weaken the resonance each other, depending on the coupling
constants. Secondly, as the
$\chi$-particles are produced significantly,
the suppression effect by the second term in Eq.~(\ref{B18})
becomes crucial.
This means that in case of non-minimal coupling, we have to
consider not only ordinary back reaction effects
to the inflaton field and metric but also
the suppression effect by this term. In what follows,  we
will  investigate these
issues in detail.

%%%%%%%%%%%%%%%%%%%%%%%%%%%%%%%%%%%%%%%%%%%%%%%%%%
%
%                                                 %
\section{The resonance by positive coupling $\xi$}

%
%                                                 %
%%%%%%%%%%%%%%%%%%%%%%%%%%%%%%%%%%%%%%%%%%%%%%%%%%

%************************************
In this section, we investigate $g=0$, $\xi>0$ case.
First, however, we briefly review the ordinary $g$-resonance,
i.e., the case with $g\ne 0$, $\xi=0$  for
comparison\cite{KLS1,KLS2}.  In this and next two sections, we
mainly study the massless
$\chi$ field.  For the massive case, we will give some
discussion in the end of each section.

If the back reaction of the $\chi$
field to the inflaton field and metric are neglected, the
inflaton field oscillates almost coherently  with damping
factor $3H$ and is approximately described as
%%%%%%%%%%%%%%%
\begin{eqnarray}
\phi  =  \Phi(t) \sin mt.
\label{P30}
\end{eqnarray}
The amplitude $\Phi(t)$ decreases with time as
\begin{eqnarray}
\Phi(t)   =  \frac{M_{\rm PL}}{\sqrt{3\pi} mt} .
\label{P3}
\end{eqnarray}
%%%%%%%%%%%%%%%
Then the time-dependent frequency of the each component
$Y_k$ becomes
%%%%%%%%%%%%%%%
\begin{eqnarray}
\omega_k^2 &=& \frac{k^2}{a^2} + g^2\phi^2
= \frac{k^2}{a^2} + g^2\Phi^2 \sin^2 mt.
\label{P3-1}
\end{eqnarray}
%%%%%%%%%%%%%%%
We can reduce Eq.~(\ref{B19}) to the well known
Mathieu equation
%%%%%%%%%%%%%%%
\begin{eqnarray}
\frac{d^2 Y_k}{d z^2} + \left[A_k -2q \cos 2z \right] Y_k=0,
 \label{P3-2}
\end{eqnarray}
%%%%%%%%%%%%%%%
where $z=mt$ and
%%%%%%%%%%%%%%%
\begin{eqnarray}
A_k= 2q + \frac{\bar{k}^2}{a^2} ,
 \label{P3-3}
\end{eqnarray}
%%%%%%%%%%%%%%%
%%%%%%%%%%%%%%%
\begin{eqnarray}
q=\frac{g^2\Phi^2}{4m^2}.
\label{P3-4}
\end{eqnarray}
%%%%%%%%%%%%%%%
$\bar{k}$ is normalized by $m$ as $\bar{k}=k/m$.
 Stability or instability with Eq.~(\ref{P3-2}) depends
on the variables of
$A_k$ and $q$, which is shown  by a
stability-instability chart (see Fig.~1)\cite{Mathieu}. In the
unstable region (the lined region in Fig.~1), $Y_k$  grows
 exponentially as $Y_k
\propto {\rm exp} (\mu_k z)$  with the Floquet index
$\mu_k$  and particles with momentum $k$ are
produced. For small $q$, the width of the instability band is
small and few $k$-modes grow by this resonance.
This is called the narrow resonance. On the other hand, for the
large
$q$, the resonance can occur for a broad range of the momentum
$k$-space. Since the
growth rate $\mu$ of the $\chi$-particle
becomes larger as the increase of the variable $q$, this
resonance gives more efficient particle production than narrow
one. This is called the broad resonance.
Note that initial amplitude $\Phi$ of the inflaton field
and coupling constant $g$ play important roles to
decide whether
resonance is narrow or broad.
In the $g$-resonance case, the allowed region on the  Mathieu
chart is determined by the Eq~(3.5) as
%%%%%%%%%%%%%%%
\begin{eqnarray}
A_k \geq 2q.
 \label{P3-5}
\end{eqnarray}
%%%%%%%%%%%%%%%
Hence the broadest resonance is given by the limit line  $A_k
=2q$.

When $q$ is sufficiently large initially, the resonance of  each
mode occurs {\it stochastically}\cite{KLS1,KLS2}. In this case,
the frequency  $\omega_k$  decreases by cosmic  expansion and
$\omega_k$ dramatically changes within each oscillation of the
inflaton field,  so the phases of $\chi$ field at successive
moment of $\phi=0$ are not correlated each other.  At the first
stage of the resonance, the fields cross large number of
instability bands. The periods when they are in the instability
band are so short that the resonance can not occur efficiently
compared with that in the Minkowski spacetime. However,
nevertheless, the number of
$\chi$-particles can still grow exponentially. As $q$ becomes
smaller, the universe expansion slows down, and the fields
stay in each resonance band for a longer time. When $q$ drops
down to about 1, the first instability band
%%%%%%%%%%%%%%%
\begin{eqnarray}
1-q-\frac18 q^2 \leq A_k \leq 1+q-\frac18 q^2,
 \label{P3-6}
\end{eqnarray}
%%%%%%%%%%%%%%%
becomes important.
When the variables decrease below the lower boundary  of
Eq.~(\ref{P3-6})  by the expansion of the
Universe, the resonance terminates. We have to note here that
there is another mechanism
which terminates the resonance.
When the initial value of $q$ is large ($g~\mbox{\raisebox{-1.ex}{$\stackrel
   {\textstyle>}{\textstyle\sim}$}}~
 3.0 \times 10^{-4}$),
$\chi$-particles are produced efficiently and  the back
reaction onto the inflaton field cannot be ignored. This makes
the oscillation of the inflaton field incoherent and finally
stops the resonance. That is called the back reaction effect
of the $\chi$ field.

%************************************

Now we turn to the case of $g=0, \xi>0$.
In this case the resonance occurs by the coupling between the
$\chi$ field and the spacetime curvature $R$\cite{BL}.
The $G(\langle\chi^2\rangle)$ term defined by Eq.~(\ref{B18})
becomes
%%%%%%%%%%%%%%%
\begin{eqnarray}
G(\langle\chi^2\rangle) &=&
   \frac{1}{(1-\eta)\left[1-(1-6\xi)\eta\right]}
   \biggl[
      2\xi\kappa^2m^2\phi^2
      -(1-3\eta)\xi\kappa^2 \dot{\phi}^2 \nonumber \\
   & &
      +\frac{\eta}{1-\eta}\frac{4
       +(1-6\xi)(1-5\eta)}{1-(1-6\xi)\eta}
      \left\{ 2\xi\kappa^2m^2\phi^2
             -(1-\eta)\xi\kappa^2\dot{\phi}^2 \right\}
         \biggr] .
\label{P1}
\end{eqnarray}
%%%%%%%%%%%%%%%
We  easily find that if $\eta$
increases up to the order of unity,  $G(\langle\chi^2\rangle)$
 diverges and the frequency of ($\ref{B20}$) increases to
infinity.
The resonance seems to continue effectively  at the final
stage of preheating. However, this is not the case.
Our numerical calculation shows that the resonance terminates
when
$\eta$ is much smaller than unity in any cases. With the
condition of $\eta
\ll1$, $G(\langle\chi^2\rangle)$ is  rewritten
as
%%%%%%%%%%%%%%%
\begin{eqnarray}
G(\langle\chi^2\rangle)
\approx \frac{\xi\kappa^2m^2}{(1+6\xi\eta)^2} \left(2\phi^2
-\frac{\dot{\phi}^2 }{m^2}\right) .
 \label{P2}
\end{eqnarray}
%%%%%%%%%%%%%%%
Note that there exists the suppression factor
$(1+6\xi\eta)^{-2}$, which makes the amplitude of the $\chi$
field small effectively. Even in the case of $\eta \ll1$, we
can expect that this suppression plays an
important role when $\xi$ takes large values.
We will see later that this effect becomes significant in the
case of
$\xi~\mbox{\raisebox{-1.ex}{$\stackrel
     {\textstyle>}{\textstyle\sim}$}}~70$.
When $\eta$ is small compared with unity, the back reaction to
the inflaton field and the metric is negligible and
we find from Eq.~($\ref{B13}$)
that the $\varphi$ field oscillates almost coherently.

By using these relations, we can rewrite the equation of $Y_k$
in the form of the Mathieu equation
%%%%%%%%%%%%%%%
\begin{eqnarray}
\frac{d^2 Y_k}{d z^2} + \left[A_k -2q \cos (2z- \alpha)\right]
Y_k=0,
 \label{P4}
\end{eqnarray}
%%%%%%%%%%%%%%%
where
%%%%%%%%%%%%%%%
\begin{eqnarray}
A_k=\frac{\bar{k}^2}{a^2} +\frac{2p}{(1+6\xi \eta)^2},
 \label{P5}
\end{eqnarray}
%%%%%%%%%%%%%%%
%%%%%%%%%%%%%%%
\begin{eqnarray}
q=\frac{\sqrt{(2p)^2
  +2 (\xi \kappa^2  \Phi^2 )^2 } }{2(1+6\xi \eta)^2},
\label{P6}
\end{eqnarray}
%%%%%%%%%%%%%%%
%%%%%%%%%%%%%%%
\begin{eqnarray}
p=\frac{\xi \kappa^2}{4}
   \left( \Phi^2 -{\Phi^{\prime}}^2 \right) ,
 \label{P7}
\end{eqnarray}
%%%%%%%%%%%%%%%
%%%%%%%%%%%%%%%
\begin{eqnarray}
\alpha = \tan^{-1} \left(\frac{ \xi \kappa^2   \Phi
\Phi^{\prime}}
             {2p+ \xi \kappa^2 \Phi^2}\right) .
\label{P8}
\end{eqnarray}
%%%%%%%%%%%%%%%
A prime denotes a derivative with respect to $z$.
Eq.~(\ref{P4}) is not exactly the Mathieu equation, because
it contains the time dependent phase term $\alpha$.
 This term is due to the existence of $\dot{\phi}^2$ term  in
Eq.~$(\ref{P2})$. If we define a new dimensionless time parameter
$\bar{t}=mt/2 \pi$, $\alpha$ is rewritten as
%%%%%%%%%%%%%%%
\begin{eqnarray}
\alpha = \tan^{-1} \left(\frac{ -4\pi \bar{t}}{12 \pi^2
 \bar{t}^2 -1}\right),
 \label{P9}
\end{eqnarray}
%%%%%%%%%%%%%%%
where we used the relation (\ref{P3}).
 Since $\bar{t}$ represents the number of the oscillation of
the inflaton field naively, $\alpha$ approaches
zero rapidly within the first oscillation of the inflaton
field.
As a result, the time dependence of $\alpha$ does not affect
the resonance process.

Although the equation of $Y_k$ has the same form as
the ordinary $g$-resonance except for the time-dependent
phase $\alpha$, properties of resonance are
not the same.
After the first oscillation of the inflaton field, setting an
initial time as $\bar{t}=1/4$ as  Kofman et al.\cite{KLS2}, we
have the condition
%%%%%%%%%%%%%%%
\begin{eqnarray}
\Phi \gg \frac{\dot{\Phi}}{m}.
\label{P10-1}
\end{eqnarray}
%%%%%%%%%%%%%%%
Then the evolution
of $A_k$ and $q$ is approximated
as:
%%%%%%%%%%%%%%%
\begin{eqnarray}
A_k & \approx &\frac{2}{3} q + \frac{\bar{k}^2}{a^2} ,
 \label{P10} \\
q & \approx  & \frac{3}{4(1+6\xi\eta)^2}
\xi\kappa^2 \Phi^2 ~~
 \approx   \frac{1}{(1+6\xi\eta)^2} \frac{\xi}
{2\pi^2 \bar{t}^2} .
 \label{P11}
\end{eqnarray}
%%%%%%%%%%%%%%%
The line described by Eq.~($\ref{P10}$) lies below the one
which is obtained
in the ordinary $g$-resonance.
As is studied in \cite{BL}, the Ricci scalar
can be replaced with $\dot{\phi}^2$ term and non-minimal
coupling provides another contribution. This is because the above
relation $(\ref{P10})$ becomes different from the relation
$(\ref{P3-3})$ of ordinary $g$-resonance.
In fact, we can easily estimate that without $\dot{\phi}^2$ term,
$\xi$ must be larger  than $10^4$ for the effective resonance,
because we have only the similar coupling term to the
$g$-resonance in that case.
Since
the width of the instability bands are thick for large $q$
(see Fig.~1),
$\xi$-resonance gives broader resonance and we may expect the
efficient $\chi$-particle production.
However, as is found from Eq.~($\ref{P11}$),
there are two factors which decrease the variable $q$. One
is the decrease of $\Phi$ by the expansion of the Universe
and the other is the suppression effect
caused by the factor $(1+6\xi\eta)^{-2}$,
which appears only in the $\xi$-resonance.
Which factor is more important depends on the parameter  $\xi$
and the efficiency of the resonance, i.e., $\eta$. When
$\xi \eta~\mbox{\raisebox{-1.ex}{$\stackrel
     {\textstyle<}{\textstyle \sim}$}}~0.1$, this suppression effect can be
neglected. Note that the back reaction effect
is negligible because $\eta$ is always small in any parameter.
Hence, it may be worth stressing  that the
$\xi$-resonance in the $g=0$ case will terminate only by
passing through the resonance band, because the coherence of
the inflaton is not broken.

By the final value of
$q$ (=$q_f$) when the resonance ends, we can estimate the time
$t_f$ when the resonance stops and the total amount
of created $\chi$-particles
$\langle \bar{\chi}^2 \rangle_f \equiv \langle \chi^2
\rangle_f /m^2$.
In the case of $q~\mbox{\raisebox{-1.ex}{$\stackrel
     {\textstyle>}{\textstyle\sim}$}}~1$, particle creation occurs
when the frequency $\omega_k$ changes non-adiabatically,
which condition is written as
%%%%%%%%%%%%%%%
\begin{eqnarray}
\omega_k ^2 < \frac{d \omega_k}{dt}.
\label{P12}
\end{eqnarray}
%%%%%%%%%%%%%%%
Neglecting the $\dot{\Phi}$  in calculation in the
$\xi\kappa^2
\dot{\phi}^2$ term in Eq.~($\ref{P2}$),
$\omega_k^2$ yields
%%%%%%%%%%%%%%%
\begin{eqnarray}
\omega_k^2 \approx
        \frac{k^2}{a^2}
           +\frac{\xi\kappa^2 m^2 }{(1+6\xi\eta)^2}
           \left( 3 \phi^2 -\Phi^2 \right),
       \label{P13}
\end{eqnarray}
%%%%%%%%%%%%%%%
and then the non-adiabatic condition ($\ref{P12}$) becomes
%%%%%%%%%%%%%%%
\begin{eqnarray}
       \frac{k^2}{a^2} & < &
      \left[ \frac{ 3\xi\kappa^2 m^2}{(1+6\xi\eta)^2}
          \phi \dot{\phi}
         -\frac{6\xi^2 \dot{\eta}\kappa^2 m^2}{(1+6\xi\eta)^3}
        \left( 3 \phi^2 -\Phi^2 \right) \right]^{2/3}
      -\frac{\xi\kappa^2 m^2}{(1+6\xi\eta)^2}
          \left( 3\phi^2 - \Phi^2 \right).
       \label{P14}
\end{eqnarray}
%%%%%%%%%%%%%%%
Let us investigate the value of $\phi$
when the r.h.s. in Eq.~($\ref{P14}$) takes
the maximum value.
Non-adiabatic amplification occurs  mostly when  $\phi$ is
passing through around the minimum of its potential
($|\phi| \ll M_{\rm PL}$),
so we can set $\dot{\phi} \approx m\Phi$ in Eq.~($\ref{P14}$).
Moreover, $\dot{\eta}$ term is negligible compared with the
former term, which is confirmed by numerical calculation.
Then, Eq.~($\ref{P14}$) can be approximately rewritten as
%%%%%%%%%%%%%%%
\begin{eqnarray}
       \frac{k^2}{a^2}  <
        \left[ \frac{3\xi\kappa^2 m^3}{(1+6\xi\eta)^2}
   \Phi\phi  \right]^{2/3}
 -\frac{\xi\kappa^2 m^2}{(1+6\xi\eta)^2}
          \left( 3\phi^2 - \Phi^2 \right).
       \label{P25}
\end{eqnarray}
%%%%%%%%%%%%%%%
Differentiating the r.h.s. of Eq~($\ref{P25}$) with respect to
$\phi$, we find that it
takes the maximal value at
%%%%%%%%%%%%%%%
\begin{eqnarray}
 \phi_{max}  =  \frac{1}{3^{3/4}} \sqrt{\frac{m\Phi}{C^{1/2}}}
  \approx  \frac{1}{2} \sqrt{\frac{m\Phi}{C^{1/2}}}
       \label{P15},
\end{eqnarray}
%%%%%%%%%%%%%%%
where $C \equiv 3\xi\kappa^2m^2/(1+6\xi\eta)^2$.
Then the maximum value
of momentum yields
%%%%%%%%%%%%%%%
\begin{eqnarray}
      \frac{\bar{k}_{max}^2}{a^2}  =
     \frac{C \Phi^2}{3m^2}
     +\left(2^{4/3}-1  \right) \frac{\sqrt{C} \Phi}{4m},
        \label{P16}
\end{eqnarray}
%%%%%%%%%%%%%%%
where $\bar{k}_{max}^2 \equiv k_{max}^2/m^2$.
As we have the relation $\sqrt{C}\Phi/m=2\sqrt{q}$
by Eq.~$(\ref{P11})$, the maximum momentum is rewritten
in terms of $q$ as
%%%%%%%%%%%%%%%
\begin{eqnarray}
      \frac{\bar{k}_{max}^2}{a^2}
 =
     \frac{4}{3} q -\left( 2^{1/3} -\frac{1}{2} \right) \sqrt{q}
    \approx  \frac{4}{3} q +\frac{3}{4} \sqrt{q}.
    \label{P18}
\end{eqnarray}
%%%%%%%%%%%%%%%
This equation gives the maximum momentum  for the resonance.
While the  resonance terminates when the variables
$A_k$ and $q$ pass the curve of
%%%%%%%%%%%%%%%
\begin{eqnarray}
 A_k \approx1-q-\frac{1}{8} q^2,
  \label{P20}
\end{eqnarray}
%%%%%%%%%%%%%%%
in the Mathieu chart \cite{Mathieu}.
Combining the relations ($\ref{P10}$), ($\ref{P18}$) and
($\ref{P20}$), we obtain  the equation with respect to $q_{f}$
 as
%%%%%%%%%%%%%%%
\begin{eqnarray}
q_{f}^2+24q_{f}+6\sqrt{q_{f}}+8=0,
\label{P40}
\end{eqnarray}
%%%%%%%%%%%%%%%
resulting
$q_{f}=0.2165 \approx 1/5$.
%%%%%%%%%%%%%%%
If we adopt the typical momentum $k_*$,
which would be defined as
\begin{eqnarray}
 k_{*} =\frac{k_{max}}{\sqrt{2}},
  \label{P19}
\end{eqnarray}
%%%%%%%%%%%%%%%
in stead of  ${k}_{max}$, we find
%%%%%%%%%%%%%%%
\begin{eqnarray}
3q_{f}^2+56q_{f}+9\sqrt{q_{f}}+24=0,
\label{P41}
\end{eqnarray}
%%%%%%%%%%%%%%%
and $q_{f}=0.3353 \approx 1/3$.

The analytic expression of Eq.~($\ref{P40}$) or ($\ref{P41}$)
is only an approximation
because the maximal momentum $k_{max}$ will be changed
when $q$ drops down to $q~\mbox{\raisebox{-1.ex}{$\stackrel
     {\textstyle<}{\textstyle \sim}$}}~1$.
However, it gives good agreement with numerical calculation.
First, we examine the evolution of $\langle \chi^2 \rangle$,
comparing the results of numerical calculation with the
analytical estimation. For $\xi~\mbox{\raisebox{-1.ex}{$\stackrel
     {\textstyle<}{\textstyle \sim}$}}~10$, numerical
calculation shows  no effective production of $\langle \chi^2
\rangle$. This means that the initial value of $q$  is so small
and it decreases so fast because of
the expansion of the Universe that the $\chi$-field does not
stay  in the instability bands for enough time.
For $10~\mbox{\raisebox{-1.ex}{$\stackrel
     {\textstyle<}{\textstyle \sim}$}}~\xi~\mbox{\raisebox{-1.ex}{$\stackrel
     {\textstyle<}{\textstyle \sim}$}}~70$, the resonance
occurs because the period during which the $\chi$ field stays in
the broad resonance band becomes longer.
For example, in the case of $\xi=50$,
the value of $\langle \chi^2 \rangle$ first increases
exponentially by the passage of time and reaches to its maximum
value
$\langle \bar{\chi}^2 \rangle_{f} =3.090\times 10^3$
at $\bar{t}_f=2.85$ (Fig.~2(a)).
Although  $\langle \bar{\chi}^2 \rangle_{f}$ is not large
enough for
the efficient preheating,
 the parametric resonance evidently occurs.
Since the maximal value of $\xi\eta$ is  $\xi\eta_{f}  \approx
1.9
\times 10^{-4}$,
we can  ignore the suppression factor
$(1+6\xi\eta)^{-2}$ in Eq.~($\ref{P11}$).
Hence the resonance is  terminated by the expansion of the
Universe. The final value of $q$ obtained by the numerical
calculation is
$q_{f} \approx 0.312$, which
is almost the same as the  analytically estimated value with
the typical momentum $k_*$. In the case that the suppression
effect is neglected, we can estimate the time when the
resonance ceases by
%%%%%%%%%%%%%%%
\begin{eqnarray}
\bar{t}_{f}  =  \sqrt{\frac{\xi}{2\pi^2q_{f}}} \approx
\sqrt{\frac{3\xi}{2\pi^2}} .
\end{eqnarray}
%%%%%%%%%%%%%%%
For the case of $\xi=50$, $\bar{t}_{f}= 2.76$, which is  close
 to the numerical value $\bar{t}_{f}= 2.85$.
After $\langle \bar{\chi}^2 \rangle$ reaches its maximum value,
it decreases monotonically. This is  the adiabatic
damping due to the expansion of the Universe.

In order to see the $\xi$-dependence of the numbers of created
particles, we depict $\langle
\bar{\chi}^2
\rangle_{f}$ in terms of   $\xi$ in Fig.~3.
It shows that although the created particle first increases as
the coupling constant $\xi$ gets large, it rather decreases
beyond a critical value $\xi_c$, which can be understood as
follows.

When $\xi$ is less than about 100, 
$\langle \bar{\chi}^2 \rangle_{f}$
increases as
the coupling constant $\xi$ gets larger.
This is just because the initial value of $q$ is larger and
then the resonance begins in the broader bands.
For 
$70~\mbox{\raisebox{-1.ex}{$\stackrel
     {\textstyle<}{\textstyle \sim}$}}~\xi~\mbox{\raisebox{-1.ex}{$\stackrel
     {\textstyle<}{\textstyle \sim}$}}~200$, the suppression factor becomes
important. The $\xi=100$ case is shown in Fig.~2(b). At the
first stage,
$\langle \bar{\chi}^2 \rangle$ increases rapidly with the
larger growth rate
than the case with $\xi=50$, but it
reaches its maximal value
$\langle \bar{\chi}^2 \rangle_{f}=9.550\times 10^5$ soon.
Since $\xi\eta_{f}=0.240$ at the maximum point,
 the suppression effect by $(1+6\xi\eta)^{-2}$ cannot be
ignored. In Fig.~3, we find  that
$\langle \bar{\chi}^2 \rangle_{f}$ is
almost flat around $\sim 10^{6}$
in the parameter range $\xi=100 \sim 200$.

For $\xi~\mbox{\raisebox{-1.ex}{$\stackrel
     {\textstyle>}{\textstyle\sim}$}}~200$, the suppression
effect by large $\xi$ is more effective than increasing of
$\langle \bar{\chi}^2 \rangle$.
At the first stage of preheating,  the growth rate of $\langle
\bar{\chi}^2
\rangle$ becomes larger
as the increase of $\xi$ (Fig.~2(c)).
 However,
the resonance soon terminates with the smaller final value
$\langle \bar{\chi}^2 \rangle_f$  than in the case of $\xi
\approx 100$ by the suppression effect.
For example, in the case with $\xi = 1000$,
$\langle \bar{\chi}^2\rangle_{f}=8.318 \times 10^4$
and $\xi\eta_{f} = 2.090$ at $\bar{t}_f=1.31$, which is much
smaller that the amount in the case with $\xi=100$.
In fact, $\langle \bar{\chi}^2 \rangle_{f}$
decreases monotonically by the suppression effect beyond
$\xi_c \sim 100$-$200$ (Fig.~3). In the case of $\xi~\mbox
{\raisebox{-1.ex}{$\stackrel
     {\textstyle>}{\textstyle\sim}$}}~200$,
we can estimate the value of
$\langle \bar{\chi}^2 \rangle_{f}$ as follows.
Assuming that the  mode with the typical momentum $k_*$
is the leading mode of the growth of
$\langle \bar{\chi}^2 \rangle$, we can rewrite Eq.~(\ref{P11})
as
%%%%%%%%%%%%%%%
\begin{eqnarray}
\langle \bar{\chi}^2 \rangle_{f} =
\frac{(M_{\rm PL}/m)^2}{48\pi\xi^2}
\left(\sqrt{\frac{3\xi}{2\pi^2\bar{t}_{f}^2}}
-1 \right),
  \label{P21}
\end{eqnarray}
%%%%%%%%%%%%%%%
at the maximal point.
In our numerical calculation, we find that
$\bar{t}_{f}$ is well approximated by the constant value 1.31
when $\xi$ is greater than 500.
Then,
$\langle \bar{\chi}^2 \rangle_{f}$ decreases as
%%%%%%%%%%%%%%%
\begin{eqnarray}
\langle \bar{\chi}^2 \rangle_{f} \propto \xi^{-3/2},
  \label{P22}
\end{eqnarray}
%%%%%%%%%%%%%%%
which is confirmed by our numerical result given in Fig.~3
for
$\xi~\mbox{\raisebox{-1.ex}{$\stackrel
     {\textstyle>}{\textstyle\sim}$}}~500$.

In Table~I, we also show our numerical results and  the
estimated values ($\ref{P21}$) for the maximal value
$\langle \bar{\chi}^2 \rangle_f$.
The present analytical estimation gives a good approximation
to  the numerical results.  A small
difference between those may be  due to the naive condition of
Eq.~($\ref{P12}$). If we take into account  all momenta larger
than $k_*$, we will  obtain the  larger values of
$\langle \bar{\chi}^2 \rangle_{analytic}$,
which estimation gives closer value to the numerical one.

Finally, we should mention the effect of a
mass of the $\chi$-field.
In general, the mass effect of the $\chi$-field works as a
suppression factor, because the relation between $A_k$ and
$q$ in
 $(\ref{P5})$ is modified as
%%%%%%%%%%%%%%%
\begin{eqnarray}
A_k=\frac{\bar{k}^2}{a^2} +\frac{2p}{(1+6\xi\eta)^2}
+\frac{m_{\chi}^2}{m^2} .
\label{PM}
\end{eqnarray}
%%%%%%%%%%%%%%%
However, since the resonance band is wider in
$\xi$-resonance compared with
$g$-resonance, the effective production of
$\chi$-particles is still expected even if the mass term is
taken into account. In fact we have found that
if the mass is  order
$10^{13}$ GeV~(namely the same order of the inflaton mass),
$\chi$-particles are
 created as same as the massless case when $\xi$ is about $100
\sim 200$. The example is given in Fig.~4. The final value of
$\langle
\bar{\chi}^2
\rangle$  in the case of $g=0, \xi=200, m_{\chi}=m$ is
about $10^6$, which is almost the same of the massless case.
However, for $\xi <1000$,  it is difficult to have a resonant
production of 
$\chi$-particles  with the mass
larger than $100 m (\sim 10^{15}$ GeV), because the
relation between
$A_k$ and $q$ deviates from the  resonance
bands due to the mass term. When $\xi$ is much larger than
1000, although it is not likely,  
we still expect production of massive $\chi$-particle
whose mass is of order $10^{15}$ GeV at the initial stage of
preheating. However, as the production of $\chi$-particle
proceeds,
$\xi\eta$ suppression becomes efficient for such a large
value of $\xi$, resulting in a small total amount of
$\langle \bar{\chi}^2 \rangle$.
Namely, massive $\chi$-particle production is possible when
$\xi$ is sufficiently large, but the final amount of
$\chi$-particle gets small by the $\xi\eta$ suppression
effect. Whether or not such a small amount of production can
still provide us the baryonsynthesis is another problem,
which we do not investigate here.

%%%%%%%%%%%%%%%%%%%%%%%%%%%%%%%%%%%%%%%%%%%%%%%%%%
%%%%%%%%%%%%
\section{The resonance by negative coupling $\xi$ }
%%%%%%%%%%%%
%%%%%%%%%%%%%%%%%%%%%%%%%%%%%%%%%%%%%%%%%%%%%%%%%%%

Next, we investigate the case of $g=0, \xi<0$.
Since $\eta$ is small compared with unity as we will  see later
by  numerical calculation,
we can use the Mathieu equation $(\ref{P4})\sim (\ref{P8})$
to analyze our numerical results
in this case as well.
Neglecting $\dot{\Phi}$ term,
the relation between $A_k$ and $q$ is now
%%%%%%%%%%%%%%%
\begin{eqnarray}
A_k \approx -\frac{2}{3} q + \frac{\bar{k}^2}{a^2} ,
 \label{N1}
\end{eqnarray}
%%%%%%%%%%%%%%%
where
%%%%%%%%%%%%%%%
\begin{eqnarray}
q \approx \frac{1}{(1+6\xi\eta)^2} \frac{|\xi|}
{2\pi^2 \bar{t}^2} .
 \label{N2}
\end{eqnarray}
%%%%%%%%%%%%%%%
Note that $A_k$ can take a negative value because of   the
negative coupling $\xi$. This fact makes the properties of
resonance  quite different from
the case of $\xi>0$ as was pointed out in
Ref.~\cite{BL}.
In the regions of $A_k<0$, a new instability
band (zeroth instability band) extends below some curve which
is approximated by $A_k \approx -q^2/2$ when
$q$ is small (see Fig.~1).
One of the important features is that this zeroth instability
band reaches $A_k=0$ in the  limit of $q=0$.
Consider the modes with very small momentum $k$.
The line of $A_k=-2q/3$  crosses to the curve
$A_k \approx -q^2/2$
at  $q \approx 1.4$.
Hence, even if $q$
evolves below
 unity by the expansion of the Universe, there remain some
unstable modes
 in this instability band.
Since the Floquet index $\mu_k$ of this instability band
behaves as $\mu_k \sim \sqrt{q}$ when $q$ is small,
resonance continues to occur until $q$ vanishes.
Moreover, most of the $A_k<0$ region is covered by either
the zeroth or higher instability bands. Almost all modes
contribute to the resonant process in almost all the time with
high growth rate. Hence
one can expect the very effective particle production.

We show numerical results for the evolution of  $\langle
\chi^2
\rangle$  in Fig.~5. For $\xi =-20$,
$q$ takes initially
 the value $\sim 16$
and modes are either in the first or in  higher instability
bands, depending on the momentum.
However, after the first oscillation of the inflaton field, $q$
decreases to $\sim 1$ and some modes enter into the zeroth
instability  band.
$\langle \chi^2 \rangle$  increases exponentially mainly by
those modes  during the first
several  oscillations of the inflaton field.
Thereafter the growth rate becomes small
and $\langle \chi^2 \rangle$  approaches the constant value
$\langle \bar{\chi}^2 \rangle_{f}=4.1 \times 10^{7}$. At this
stage, the  production of $\chi$-particle balances with the
dilution  by the Hubble expansion.
For $\xi =-100$, $\langle \chi^2 \rangle$ increases more
rapidly than the case of $\xi = -20$ at the first stage
and reaches to its maximum value
$\langle \bar{\chi}^2 \rangle_{f}=4.4 \times 10^{6}$
at $\bar{t}=7.7$.
The main growth of $\langle \chi^2 \rangle$ occurs until
$\bar{t}=2$, when $q$ is of order 1.
After $\bar{t}=2$, the $\chi$-field  enters the zeroth
instability band, and  the increase of $\langle
\chi^2 \rangle$  stops at $\bar{t}=7.7$.
After that, $\langle \chi^2 \rangle$ takes almost a constant
value. This behavior is universal for the case with  $\xi<0$.
To understand the present results more deeply,
we shall estimate the value of $\langle \bar{\chi}^2
\rangle_{f}$.

%%%%%%%%%%%%%%%
{}From the relations ($\ref{B8}$) and ($\ref{B7}$),
we can obtain the following relation:
%%%%%%%%%%%%%%%
\begin{eqnarray}
\frac{d}{dt} \langle\chi^2\rangle
  & = &
    \sqrt{\frac{\langle \chi^2 \rangle}{\langle F^2\rangle\langle
X^2 \rangle}}
   \frac{1}{a^3} \left( \frac{d}{dt} \langle Y^2 \rangle
  -3H \langle Y^2 \rangle \right), \nonumber
  \\
   & = & \sqrt{\frac{\langle \chi^2 \rangle \langle X^2 \rangle }
   {\langle F^2\rangle}    }
   m \left( 2\mu-3\bar{H} \right),
   \label{N4}
\end{eqnarray}
%%%%%%%%%%%%%%%
where we have used the expression $\langle Y^2 \rangle =
a^3\langle X^2 \rangle \sim e^{2\mu mt}$  with $\mu$ being the
Floquet index and
$\bar{H}=H/m$.
The growth of $\langle \chi^2 \rangle$ stops when the r.h.s.
vanishes, i.e., when the dilution effect by expansion
of the Universe surpasses the particle creation rate.
Neglecting the back reaction effect on the metric,
the evolution of the Hubble parameter is approximately written
by
%%%%%%%%%%%%%%%
\begin{eqnarray}
    \bar{H }\approx \sqrt{\frac{4\pi}{3}} \frac{\Phi}{M_{\rm PL}}
       \approx \frac{1}{3\pi \bar{t}} .
   \label{N5}
\end{eqnarray}
%%%%%%%%%%%%%%%
While, the relation between $\mu$ and $A$
in the zeroth instability band
can be written
by \cite{Mathieu}
%%%%%%%%%%%%%%%
\begin{eqnarray}
\mu \approx \left[ -A+\frac{(A-1)q^2}{2(A-1)^2-q^2}
\right]^{1/2}.
 \label{N7}
\end{eqnarray}
%%%%%%%%%%%%%%%
Considering the modes which are close to the line of
$A=-2q/3 $, the Floquet index $\mu$  is given as
%%%%%%%%%%%%%%%
\begin{eqnarray}
\mu  \approx  \sqrt{\frac{2q}{3} }
\approx   \frac{1}{1+6\xi\eta} \sqrt{\frac{|\xi|}{3}}
\frac{1}{\pi \bar{t}} .
\label{N9}
\end{eqnarray}
%%%%%%%%%%%%%%%
Until $\langle \chi^2 \rangle$ reaches to its maximum value,
$\mu$ decreases faster than $1/\bar{t}$ by the existence of the
suppression
factor $1/(1+6\xi\eta)$. Because of this behavior,
the Hubble expansion, which decreases as $\sim 1/\bar{t}$,
will catch up with the inflaton decay,
and then the preheating is terminated.
Namely, the main factor which stops the growth of $\langle
\chi^2 \rangle$ is the production of $\langle \chi^2
\rangle$ itself.
Substituting Eqs.~($\ref{N5}$), ($\ref{N9}$) to Eq.~($\ref{N4}$),
the growth rate of
$\langle \chi^2 \rangle$ is approximately given as
%%%%%%%%%%%%%%%
\begin{eqnarray}
\frac{d}{dt} \langle\chi^2\rangle
  \approx \sqrt{\frac{\langle \chi^2 \rangle \langle X^2
\rangle}
  {\langle F^2\rangle}}
    \frac{m}{\pi \bar{t}} \left[
   \frac{1}{1+6\xi\eta} \sqrt{\frac{4|\xi|}{3}} -1 \right].
\label{N10}
\end{eqnarray}
%%%%%%%%%%%%%%%
When the term in the square bracket vanishes, the growth of
$\langle \chi^2 \rangle$ ceases.
By this condition we find the final value of
$\langle \chi^2 \rangle$ as
%%%%%%%%%%%%%%%
\begin{eqnarray}
\langle \bar{\chi}^2 \rangle_{f}  =\frac{(M_{\rm
PL}/m)^2}{48\pi\xi^2}
\left[ \sqrt{\frac{4|\xi|}{3}} -1 \right].
\label{N11}
\end{eqnarray}
%%%%%%%%%%%%%%%
Note that the final value of $\langle \chi^2 \rangle$
depends on only $\xi$,
but not on $\bar{t}_{f}$.
Once $\langle \chi^2 \rangle$ reaches to its maximum,
$\mu$ and $H$
both decreases as $1/\bar{t}$.
Note that in the case of $g=0, \xi>0$, the dependence of $\mu$
is  $\mu \sim q \sim 1/\bar{t}^2$ when $q$ is close to zero
and that $\langle \chi^2 \rangle$ does not approach to constant
value but adiabatically decreases by the Hubble expansion.

{}From Eq.~(\ref{N10}), we find that the
resonance does not occur in the
case of $0<|\xi|<0.75$, and this is confirmed by
numerical calculation.
In this parameter range, the growth rate $\mu$ is small
as compared with the Hubble expansion rate.
By Eq.~$(\ref{N11})$, we guess that $\langle \bar{\chi}^2
\rangle_{f}$ may take the maximal value at $\xi
\approx -1.33$.  However, this is not the case actually.
In $-3~\mbox{\raisebox{-1.ex}{$\stackrel{\textstyle<}{\textstyle\sim}$}}~\xi~\mbox{\raisebox{-1.ex}{$\stackrel{\textstyle<}{\textstyle\sim}$}}~-1$ case, since the creation rate of $\chi$-particle
is small, it takes much time to complete the $\chi$-particle
production.
As time passes, a contribution of produced $\chi$-particles to the
Hubble expansion rate becomes comparable to that of the inflaton
field, and the estimation $(\ref{N11})$ can not be applied.
The growth of $\langle\bar{\chi}^2\rangle$ stops
before $\xi\eta$ suppression effect in Eq.~(\ref{N10})
becomes significant.
Hence for $-3~\mbox{\raisebox{-1.ex}{$\stackrel{\textstyle<}{\textstyle\sim}$}}~\xi~\mbox{\raisebox{-1.ex}{$\stackrel{\textstyle<}{\textstyle\sim}$}}~-1$, although the $\chi$-particle production 
is possible, the final abundance of $\chi$-particle is not so large.
For example, in $\xi=-2$ case, the numerical value of 
the final abundance is $\langle\bar{\chi}^2\rangle_f=
3.2 \times 10^3$, which is much smaller than the estimated value
$\langle\bar{\chi}^2\rangle_f=1.0 \times 10^9$  by 
Eq.~(\ref{N11}).
On the other hand, for $\xi~\mbox{\raisebox{-1.ex}{$\stackrel{\textstyle<}{\textstyle\sim}$}}~-3$, $\xi\eta$ suppression effect
plays a crucial role to teriminate the increase of 
$\langle\bar{\chi}^2\rangle$.
Numerically, $\langle\bar{\chi}^2\rangle_f$ takes maximal
value $\langle\bar{\chi}^2\rangle_{max} \sim 3.2 \times 10^8$
at $\bar{t}_f \sim 3.3 \times 10^5$ when $\xi \sim -4$.
In this case, although the growth rate is still small
compared with other cases as $g>0, \xi=0$ and $g=0, \xi>0$,
the final fluctuation is large.
For $\xi~\mbox{\raisebox{-1.ex}{$\stackrel{\textstyle<}{\textstyle\sim}$}}~-4$,  
$\langle \bar{\chi}^2 \rangle$ continues to grow
until $\chi$-particles are  significantly produced,
and finally $\xi\eta$ suppression effect terminates the resonance.
In these cases, the analytic estimation based on
Eq.~$(\ref{N11})$ can be applied.
By Eq.~$(\ref{N11})$, we expect that 
$\langle \bar{\chi}^2 \rangle_f$  decreases as the decrease
of $\xi$ ($\mbox{\raisebox{-1.ex}{$\stackrel{\textstyle<}{\textstyle\sim}$}}~-4$).
For example, numerical values of the final fluctuation are $\langle\bar{\chi}^2\rangle_f=1.122 \times 10^8$ for $\xi=-10$ and $\langle\bar{\chi}^2\rangle_f=4.130 \times 10^7$ for $\xi=-20$.
In $\xi=-10$ case,  $\bar{t}_f$ is about $\bar{t}_f \sim
1.0 \times 10^3$ and the growth rate is $\mu \sim$ 0.001.
For $\xi=-20$, the growth rate increases up to $\mu \sim 0.1$,
and the resonance ends at $\bar{t}_f=28.28$.
In the case of $|\xi|\gg 1$, the resonant time becomes
very short by this suppression and the final value of
$\langle \bar{\chi}^2 \rangle$ is reduced. 
Eq.~(\ref{N11})  shows that
$\langle \bar{\chi}^2 \rangle_{f}$ decreases as $|\xi|^{-3/2}$
for $|\xi|\gg 1$.

In TABLE~II, we show the analytically estimated and numerical
values of $\langle\bar{\chi}^2\rangle_{f}$ for various cases.
We find that the analytical estimation gives  good
agreement for $\xi~\mbox{\raisebox{-1.ex}{$\stackrel{\textstyle<}{\textstyle\sim}$}}~-4$. The small discrepancy comes from the fact that
we mainly considered the  modes  close to $k=0$, which gives
the largest contribution in most of the stages.
Also, since the actual
Hubble parameter is larger than is estimated by Eq.~(\ref{N5}), 
this decreases the estimated value of 
$\langle \bar{\chi}^2 \rangle_{f}$.
Note that the final value of $\langle \bar{\chi}^2
\rangle_{f}=3.162 \times 10^8 $ at $\xi=-4$ is larger than the maximal value  in the case of $g=0, \xi> 0$ ($\langle \bar{\chi}^2 \rangle_{max} \sim 10^6$ at $\xi=100 \sim 200$).
Since the resonance bands of negative coupling are broader
than those of positive coupling in previous section,
the more $\chi$-particles are created and the
final value of $\langle \bar{\chi}^2 \rangle$
becomes larger.
However, when the value of $|\xi|$ becomes large,
$\langle \bar{\chi}^2 \rangle_{f}$ decreases as
$|\xi|^{-3/2}$, which is the same as positive coupling case.
We  conclude that the suppression effect
controls the finial value of $\langle \bar{\chi}^2 \rangle$
for large $|\xi|$ in $\xi$-resonance case.

When the mass of $\chi$-field is taken into account,
the resonance is suppressed as the case of $g=0, \xi>0$.
However, since the resonance band is broader than those of
$g=0, \xi>0$ case, we expect considerable production of
$\chi$-particles for $\xi<0$.
One important property 
in negative $\xi$ case is that massive 
$\chi$-particle is hard to be created
once $\chi$-field enters the zeroth instability band.
This is because $\chi$-field deviates from the zeroth 
instability region by the mass effect.
Even in the case of $\xi=-20 \sim -30$ when massless 
$\chi$-particles are effectively produced as 
$\langle \bar{\chi}^2 \rangle_f~\mbox{\raisebox{-1.ex}{$\stackrel{\textstyle>}{\textstyle\sim}$}}~5 \times 10^7$, massive 
$\chi$-particle production is strongly suppressed because
$\chi$-field deviates from the zeroth instability band
soon after the first oscillation of the inflaton field. 
For example, in the case of $g=0, \xi=-30, m_{\chi}=m$, we find
$\langle \bar{\chi}^2 \rangle_{f} \sim 10^3$, which is not an
effective resonance. 
Rather, in the case of $g=0, \xi=-100 \sim -200$, in spite of the $\xi\eta$ suppression effect, 
the produced $\chi$-particles exceed more than 
$\langle \bar{\chi}^2 \rangle_{f}=10^6$ 
for $m_{\chi}=m$ (Fig.~6).
However, if $m_{\chi}~\mbox{\raisebox{-1.ex}{$\stackrel{\textstyle>}{\textstyle\sim}$}}~10m$,  it becomes difficult to produce more than
$\langle \bar{\chi}^2 \rangle_{f} \sim 10^6$ even for
$\xi=-100 \sim -200$.
We show in Fig.~7 the final abundance of $\chi$-particle
as a function of $m_{\chi}$ for $\xi=-1000$ case.
In this case, we find that $\chi$-particle whose mass is more 
than $m_{\chi} \sim 100m$ can not be created.
In order to produce the GUT scale particles
($m_{\chi} \sim 10^3 m \sim 10^{16}$ GeV),
the value of $A_k$ is more than $A_k =10^6$
since $m_{\chi}^2/m^2 \sim10^6$ in Eq.~(\ref{PM}).
This means that the resonance does not happen at all from
the very beginning unless $\xi~\mbox{\raisebox{-1.ex}{$\stackrel
 {\textstyle<}{\textstyle \sim}$}}~-10^6$ [Note $q \approx |\xi|$
initially]. Moreover, when $|\xi|$ is extremely large
such as $\xi \sim -10^6$, $\chi$-particle  with the
mass of order
$10^{16}$ GeV can  be created initially,  but the final
amount will be largely reduced  by the $\xi\eta$
suppression effect. 
As is the same with $g=0,\xi>0$ case, whether or not such a small
amount of production can provide us the baryonsynthesis is
another problem.
We then conclude that for massive case,
effective $\chi$-particle production is expected when the $\chi$
field does not deviate from instability bands initially and
$\xi\eta$ suppression effect is not too strong. 

%%%%%%%%%%%%%%%%%%%%%%%%%%%%%%%%%%%%%%%%%%%%%%%%%%
%%%%%%%%%%
%                                                          %
\section{The combined resonance}    %
%                                                          %
%%%%%%%%%%%%%%%%%%%%%%%%%%%%%%%%%%%%%%%%%%%%%%%%%%
%%%%%%%%%%

In previous two sections, we have shown that the non-minimal
coupling will give rise to the parametric resonance depending
on the value of coupling constant $\xi$.
Here we shall  investigate a combination of
$g$-resonance
 and $\xi$-resonance (the combined resonance).
When $g\ne 0$, $g^2 \langle\chi^2 \rangle$ term in the equation
for the inflaton field (\ref{B13}) becomes important
at the final stage of the evolution and changes the
effective mass of the inflaton field. By this back reaction
effect, the production of $\chi$-particles is suppressed.
Similarly, produced
$\chi$-particles will affect the metric by the coupling between
the $\chi$ and  inflaton fields.
However, first, in order to  examine the naive
structure of the present system,
we shall rewrite the equation for the $\chi$ field under the
condition of
$g^2 \langle\chi^2 \rangle \ll m^2$ and $\eta \ll 1$.
With  Eqs.~(\ref{P30}) and (\ref{P3}),
the equation for the $\chi$ field
is reduced to the form of Mathieu equation (\ref{P4}) with
(\ref{P5}), (\ref{P8}), and $q$ and $p$ being defined as
%%%%%%%%%%%%%%%
\begin{eqnarray}
q=\frac{\sqrt{(2p)^2
  +(g^2\Phi^2/m^2+2\xi\kappa^2\Phi^2) \xi \kappa^2 \Phi^2 }}
    {2(1+6\xi \eta)^2},
\label{C3}
\end{eqnarray}
%%%%%%%%%%%%%%%
%%%%%%%%%%%%%%%
\begin{eqnarray}
p=\frac14
   \left( \frac{g^2\Phi^2}{m^2}+\xi\kappa^2\Phi^2
 -\xi\kappa^2{\Phi^{\prime}}^2 \right) .
 \label{C4}
\end{eqnarray}
%%%%%%%%%%%%%%%
With the condition ($\ref{P10-1}$), $A_k$ and $q$ are rewritten
as
%%%%%%%%%%%%%%%
\begin{eqnarray}
          A_k  & \approx & \frac{\bar{k}^2}{a^2}
          +2q \frac{g^2 \Phi^2/m^2 +\xi \kappa^2 \Phi^2}
          {\left| g^2 \Phi^2/m^2
           +3\xi \kappa^2 \Phi^2 \right|}
            \approx  \frac{\bar{k}^2}{a^2}
          +2q \frac{g^2 +8 \pi \xi (m/M_{\rm PL})^2 }
          {\left|g^2 +24 \pi \xi (m/M_{\rm PL})^2 \right|} ,
          \label{C6}
\\
 q   & \approx &
         \frac{1}{4(1+6\xi \eta)^2} \left| \frac{g^2 \Phi^2}{m^2}
           +3\xi \kappa^2 \Phi^2 \right|
           \approx
         \frac{(M_{\rm PL}/m)^2}{48\pi^3(1+6\xi\eta)^2 \bar{t}^2}
        \left| g^2 +24 \pi \xi \Bigl(\frac{m}{M_{\rm PL}}
         \Bigr)^2\right| .
    \label{C7}
\end{eqnarray}
%%%%%%%%%%%%%%%

On the Mathieu chart, there are two quantities which determine
the efficiency of the resonance. One  is the initial value
$q_i$. This determines the Floquet index, i.e., the growth
rate of the
$\chi$-particle. It becomes large as $g$ and $\xi$ get
large as seen from Eq.~(\ref{C7}). In general, if we can
neglect the back reaction effect, we expect more efficient
resonance for large
$q_i$. Another quantity is the gradient of the line on which
the variables $A_k$ and $q$ trace during evolution of the
system. It  determines how many modes  contribute
to resonance. In the present case,
the gradient is given by the ratio of $\xi$ to $g^2$. In the
minimal coupling case, $\xi =0$, so the typical line is
$A_k=2q$. In the non-minimal case, however, there are several
types of lines which show different behaviors.
Then  we shall first classify the Mathieu chart into several
regions by  changing   $\xi/g^2$
 from $\infty$ to $-\infty$ (see Fig.~8).
 When $\xi/g^2=\infty$, the resonance is the similar  to the
positive
$\xi$-resonance  discussed in Sec. III.
When $0<\xi/g^2<\infty$, the resonance occurs below the
$A_k=2q$ line.
We call the resonance in this parameter region ``new-broad
resonance".
When $\xi=0$, resonance occurs only by $g$-coupling:~$A_k=2q+
\bar{k}^2/a^2$ and the ordinary $g$-resonance is recovered.
If $\xi$ becomes negative, $A_k$ and $q$
enter the region of $A_k>2q$, and we call the resonance in  this
parameter region ``ordinary-broad resonance". When
$\xi/g^2=-(M_{\rm PL}/m)^2/24\pi$, $q$ vanishes and no resonance
occurs. As $\xi/g^2$ decreases further, the gradient gets
smaller and the resonance becomes broader again.
When $\xi/g^2=-(M_{\rm PL}/m)^2/16\pi$,
the typical line is $A_k=2q$, which is same as the
ordinary $g$-resonance in spite of the existence of
$\xi$-coupling. The parameter range
$-(M_{\rm PL}/m)^2/12\pi <\xi/g^2 < -(M_{\rm PL}/m)^2/16\pi$,
corresponds to the region between $A_k=2q$ and $A_k=2q/3$,
and the system enters the new-broad resonance region again.
As $\xi/g^2$ gets smaller further, $A_k$ and $q$ enter the
region of
$A_k< 2q/3$. We call the resonance in this region
``wide resonance".
When $\xi/g^2<-(M_{\rm PL}/m)^2/8\pi$, the gradient
becomes  negative. $\xi/g^2=-\infty$
corresponds to
the negative $\xi$-resonance discussed in the previous section.
In what follows, we will examine each cases one by one in
detail paying attention to which  parameter range in
$\xi$-coupling assists the $ g$-resonance.

Before proceeding the individual investigation,
we shall give some criteria for the back reaction effect
on the inflaton field and for the suppression effect by
$\xi\eta$ term.
\begin{enumerate}
 \item  For the back reaction effect on the inflaton field,
we shall adopt the criterion
%%%%%%%%%%%%%%%
\begin{eqnarray}
    g^2 \langle \bar{\chi}^2 \rangle ~\mbox{\raisebox{-1.ex}{$\stackrel
     {\textstyle<}{\textstyle \sim}$}}~ 0.1 .
    \label{C8}
\end{eqnarray}
%%%%%%%%%%%%%%%
If the $\langle \bar{\chi}^2 \rangle$ exceeds this criterion,
the back reaction becomes important.
In the case of $g \ne 0, \xi=0$, our numerical analysis shows
that  this  criterion corresponds to the condition of
$g ~\mbox{\raisebox{-1.ex}{$\stackrel
     {\textstyle<}{\textstyle \sim}$}}~ 3 \times 10^{-4}$.
\item  For the $\xi\eta$ suppression effect, we adopt the
criterion
%%%%%%%%%%%%%%%
\begin{eqnarray}
    \xi\eta ~\mbox{\raisebox{-1.ex}{$\stackrel
     {\textstyle<}{\textstyle \sim}$}}~0.1 .
    \label{C9}
\end{eqnarray}
%%%%%%%%%%%%%%%
\end{enumerate}
We will see that those effects are crucial for the effective
$\chi$-particle  creation at the final stage of preheating.

%%%%%%%%%%%%%%%%%%%%%%%%%%%
\subsection{$\xi>0$ (new-broad resonance)}    %
%%%%%%%%%%%%%%%%%%%%%%%%%%%

This case corresponds the parameter range,
\begin{equation}
    \frac{2}{3} q + \frac{\bar{k}^2}{a^2} <A_k < 2q
    + \frac{\bar{k}^2}{a^2},
\end{equation}
on the Mathieu chart (Fig. 8(a))
and the resonance band is broad  compared with usual
$g$-resonance case. We shall discuss the following three cases
separately.

%%%%%%%%%%%%%%%%%%%%%%%%%%%
\subsubsection{$g \ll 3 \times 10^{-4}$ case}

In the  case of $\xi=0$, i.e. the ordinary $g$-resonance, the
width of the instability bands are small when $g \ll 3 \times
10^{-4}$. The $\chi$-particles are actually created only in the
first instability band, and the resonance occurs in the narrow
band from the  beginning. As a result, $\chi$-particle is not
 produced efficiently. By taking $\xi$-coupling into account,
however, the resonance is assisted since  initial value of $q$
becomes larger  and the resonance band becomes broader as
seen from Eqs.~(\ref{C6}) and (\ref{C7}). In order to obtain
$\langle
\bar{\chi}^2 \rangle_{f} \sim10^6$,  we need   the coupling
constant $\xi=100 \sim 200$. Because the resonance occurs
mainly by $\xi$-coupling, the properties of resonance is
almost the same as $g=0, \xi>0$ case. For fixed value $\xi$,
we  find that
$g$ makes the resonance band narrower and the resonance less
effective, although it makes the initial value of $q$ larger.
In fact, the numerical calculation shows that
$\langle \bar{\chi}^2 \rangle_{f}$ with $g$ is smaller than
that without $g$ and its growth rate of
$\langle \bar{\chi}^2 \rangle$ is
 also smaller.
In this sense, $g$-coupling
weakens the $\xi$-resonance.
In this parameter region, the back reaction effect on the
inflaton mass is negligible and the suppression by $\xi\eta$
term terminates the  $\chi$-particle production.

%%%%%%%%%%%%%%%%%%%%%%%%%%%
\subsubsection{$g \sim 3 \times10^{-4}$ case}

In this case, we  find from Fig.~9 that the resonance is
divided into two stages as in the ordinary $g$-resonance,
namely first broad resonance regime and following narrow
resonance regime. We put the suffix ``$b$'' to the variables
just after the broad resonance regime as $\bar{t}_{b}$ and
$q_{b}$. Although the resonance  starts from the broad
resonance region even in the
$\xi=0$ case,  the resonance band becomes broader by adding
the $\xi$-coupling.
Furthermore, the initial value of $q$
becomes large as seen in Eq.~(\ref{C7}). Since  more
modes   contribute to the resonance, both the growth rate
of $\langle \bar{\chi}^2 \rangle$ and the value of
$\langle \bar{\chi}^2 \rangle_{b}$ get larger (see Fig.~9 and
Table III). This means that the $\xi$-coupling supports the
$g$-resonance well in the broad resonance regime. After the
resonance enters the narrow resonance regime ($\bar{t} \approx
8$), the growth rate hardly depends on $\xi$. At the
final stage, however, we cannot neglect the suppression effect
by the
$\xi \eta$ term.  As $\xi$ becomes large, the variables $A_k$
and
$q$ cross the lower boundary of the first instability band at
an earlier time and the final value of $\langle \bar{\chi}^2
\rangle_{f}$ is suppressed a little compared with
that in the ordinary $g$-resonance.
In this case, the back reaction effect is marginally less
important.

%%%%%%%%%%%%%%%%%%%%%%%%%%%
\subsubsection{$g >3 \times10^{-4}$ case}

In this parameter region, the back reaction to the inflaton field
should be taken into consideration. In the case of $\xi=0$,
$\chi$-particles are produced quite effectively in the broad
resonance regimes and $\langle \bar{\chi}^2 \rangle_{f}$
takes maximal value
$\langle \bar{\chi}^2 \rangle_{max} \approx 5.0 \times 10^7$
for $g \approx 1 \times 10^{-3}$, in which case the initial
value of
$q$ ($q_{i} \approx 1.075 \times 10^4$) is large enough  as
estimated by Eq.~(\ref{C7}).
As $g$ increases, the back reaction term
$g^2 \langle \bar{\chi}^2 \rangle$ gives significant effect
even small value  of $\langle \bar{\chi}^2 \rangle$ and
$\langle \bar{\chi}^2 \rangle_f$ rather decreases.
Even if we take $\xi$-coupling into account, it does not change
the growth rate at the first stage of the resonance
very much when $g$ is large (see Fig.~10).
For example, in the case of $g =1.0 \times 10^3$
and $\xi = 100$, the initial value estimated by Eq.~(\ref{C7})
is
$q_{i} \approx 1.083 \times 10^4$. This is almost same as
the case of $\xi=0$.  Furthermore, the relation between
$A_k$ and
$q$ is
$A_k \approx
1.99q+\bar{k}^2/a^2$,
which shows also the same broadness. Hence the
initial stage of the resonance is governed by the $g$-coupling.
In order to find the effect of the $\xi$-coupling, we may set
$\xi$ very large
($\xi~\mbox{\raisebox{-1.ex}{$\stackrel
     {\textstyle>}{\textstyle\sim}$}}~10^4$ for $g=1 \times 10^{-3}$). Such a large $\xi$
coupling, however, causes extremely strong suppression  and we
do not expect the efficient resonance.
In the final stage, either the suppression effect by $\xi\eta$
or the back reaction effect terminates the resonance depending
on the coupling
 constants $g$ and $\xi$. Generally speaking, when $\xi$ takes
a large  value, the suppression factor becomes important before
the created $\chi$-particles  cause the significant back
reaction. By two criteria (\ref{C8}) and  (\ref{C9}), we can
estimate a rough dependence. Once the created particles exceed
one of the criteria first, the resonance will terminate. Since
the r.h.s. of these criteria are  the same,
 comparing the l.h.s. terms, two effects become equivalent
when
%%%%%%%%%%%
\begin{equation}
g^2=8\pi \left(\frac{m}{M_{\rm PL}}\right)^2 \xi^2.
\end{equation}
%%%%%%%%%%%
Hence, if $g^2>8\pi (m/M_{\rm PL})^2 \xi^2$, the back reaction
terminates the resonance. Otherwise, the resonance is
terminated by the suppression effect. This estimation has been
confirmed by the numerical calculations.
As a result, it is difficult to create $\chi$-particles more
than
$\langle \bar{\chi}^2 \rangle_f \sim 5.0 \times 10^7$
even if $\xi$-coupling is taken into account.

%%%%%%%%%%%%%%%%%%%%%%%%%%%
\subsection{$- ( M_{\rm PL}/m)^2/16\pi <\xi/g^2 <0$
(ordinary-broad resonance)}    %
%%%%%%%%%%%%%%%%%%%%%%%%%%%

As we can see by Eq.~(\ref{C6}), both terms of $g$ and $\xi$
suppress the instability in this case. The gradient
of the $q$-$A_k$ line becomes steep, and
this parameter range corresponds
to the region between the line of $A_k = 2q +\bar{k}^2/a^2$
and $A_k$ axis in the Mathieu chart. This means that the
resonance  occurs in narrower bands
than that in the  case of $\xi=0$.
In particular, for $\xi/g^2=-( M_{\rm PL}/m)^2/24\pi$,
the variable $q$ vanishes. Since
the width of any instability bands vanishes for $A_k>0$,
the resonance does not occur at all although both
$g$-coupling and $\xi$-coupling exist.
As for the initial value of $q$, both terms of $g$  and $\xi$
 suppress it as seen from Eq.~(\ref{C7}). As a result
$\xi$-coupling always suppresses
$g$-resonance in any parameter range.
This effect is remarkable when $q$ is small.
However, for large
$g$, $\xi$ terms is negligible because both the gradient
of the $q$-$A_k$ line and $q_i$ depend on square of
$g$ while they depend linearly on $\xi$. Hence the properties
of the resonance are almost the same as those in ordinary
$g$-resonance if  $g>3.0 \times 10^{-4}$ and $0<|\xi|
<100$.

%%%%%%%%%%%%%%%%%%%%%%%%%%%
\subsection{$-(M_{\rm PL}/m)^2/12\pi  <\xi/g^2 <- (M_{\rm
PL}/m)^2/16\pi$ (new-broad resonance)}
%
%%%%%%%%%%%%%%%%%%%%%%%%%%%

 This parameter range corresponds to the regions between
 $A_k =2q+\bar{k}^2/a^2$
and $A_k =2q/3 +\bar{k}^2/a^2$ on the Mathieu chart (Fig.~8(b)). 
The resonance gives the same broadness as the case
discussed in the subsection~A. However, there is a different
point in an  efficiency of the resonance.
In the present case, $q$ evolves in the range of
%%%%%%%%%%%%%%%
\begin{eqnarray}
\frac{1}{(1+6\xi \eta)^2} \frac{|\xi|}{6\pi^2 \bar{t}^2} <q<
\frac{1}{(1+6\xi \eta)^2} \frac{|\xi|}{4\pi^2 \bar{t}^2} .
\label{C10}
\end{eqnarray}
%%%%%%%%%%%%%%%
Comparing this with the Eq.~(\ref{P11})  in the case
of $g=0,\xi>0$,  $q$ in Eq.~(\ref{C10}) is smaller
than  that in the case of $g=0$ by
factor 2-3 for the fixed value of $|\xi|$.
Remembering that for $g=0,\xi>0$
the value of $\xi$ must be $\xi~\mbox{\raisebox{-1.ex}{$\stackrel
     {\textstyle>}{\textstyle\sim}$}}~100$
to achieve $\langle \bar{\chi}^2
\rangle_{f} \sim 10^6$,
we can estimate that $|\xi|$ is
needed at least $|\xi|~\mbox{\raisebox{-1.ex}{$\stackrel
     {\textstyle>}{\textstyle\sim}$}}~200$ for the efficient resonance
 in the case of $\xi/g^2\approx -(M_{\rm PL}/m)^2/ 12\pi$.
This means that the parameters which cause the effective
resonance are a little more restrictive than  the
$\xi$-resonance discussed in Sec.~III. In the case of
$\xi/g^2\approx -( M_{\rm PL}/m)^2/16\pi$,
$q$ evolves on the line $A_k \approx 2q+\bar{k}^2/a^2$ and
the broadness of the resonance
bands is similar to the ordinary $g$-resonance.
For this reason, we need the value of $|\xi|$ that is more than
10000 for the effective resonance, but such a large $|\xi|$
causes the
 strong $\xi\eta$ suppression and the final value of  $\langle
\chi^2 \rangle$
 is reduced.
We show the numerical results on the $A_k=2q/3$ line in Table~IV,
have confirmed the above analysis.
If $|\xi|$ is less
than 100, the resonance hardly occurs. For example, when $\xi
=-100$ and
$g=6.140 \times 10^{-5}$, the final value of
$\langle \bar{\chi}^2 \rangle$ is
$\langle \bar{\chi}^2 \rangle_{f} =5.0 \times 10$.
On the other hand, in the case of $\xi =-200$ and $g=7.089
\times 10^{-5}$,  the final value is $\langle
\bar{\chi}^2\rangle_{f} =1.0 \times10^{5}$.
 $g$-resonance is in fact assisted by $\xi$-coupling in  this
case. However, it is difficult to produce the $\chi$-particles
more than  $\langle
\bar{\chi}^2\rangle_{f} \sim 10^6$ in this parameter range.

%%%%%%%%%%%%%%%%%%%%%%%%%%%
\subsection{$-\infty <\xi/g^2<-( M_{\rm PL}/m)^2/12\pi$
(wide resonance)}    %
%%%%%%%%%%%%%%%%%%%%%%%%%%%

This parameter range corresponds to the regions between
$A_k =2q/3 +\bar{k}^2/a^2$ and $A_k=-2q/3 +\bar{k}^2/a^2$
(Fig.~8(b)). Since the contribution from the negative $\xi$-coupling
surpasses the $g$-coupling,
the resonance band becomes quite broad as $\xi/g^2$ decreases.
For $-( M_{\rm PL}/m)^2/8\pi <\xi/g^2<-( M_{\rm PL}/m)^2/12\pi$,
the gradient of the $q$-$A_k$ line is positive and the
resonance terminates when the variables $q$ and $A_k$ cross the
lower boundary of the first
instability band as the former cases. After the resonance
terminates, 
$\langle \bar{\chi}^2 \rangle$ decreases gradually by the
adiabatic  expansion. We show the result of the numerical
calculations in the $\xi/g^2 = -( M_{\rm PL}/m)^2/8\pi$ case,
which correspond to
$A_k=0$,  in Table~V. In the case of $|\xi|~\mbox{\raisebox{-1.ex}{$\stackrel
     {\textstyle<}{\textstyle \sim}$}}~50$,
resonance does not occur because the initial value of $q$ is
too small. However as $|\xi|$ becomes large, $\langle
\bar{\chi}^2
\rangle_{f}$ increases and for $\xi=-200$ and $g=7.089 \times
10^{-5}$, $\langle \bar{\chi}^2
\rangle_{f} =1.0 \times10^{6}$. {}From these results we can
conclude that resonance is more effective than
the previous case.

When $\xi/g^2~\mbox{\raisebox{-1.ex}{$\stackrel
     {\textstyle<}{\textstyle \sim}$}}~-( M_{\rm PL}/m)^2/8\pi$, the $q$-$A_k$ line
passes  through the zeroth instability band near its
boundary. Since the Floquet index
is small there, it takes long time to reach the finial value of
$\langle \bar{\chi}^2 \rangle$ even for the large $|\xi|$.
$\langle \bar{\chi}^2 \rangle_{f}$  is determined by the
balance between the  creation rate
of the $\chi$-particle and the expansion rate of
the Universe (Eq.~(\ref{N11})).
As $\xi/g^2$ approaches to $-\infty$, the properties of the
resonance are almost the same as the case of $g=0, \xi<0$,  and
then
$\xi$-coupling  assists $g$-resonance. Since the existence of
$g$-coupling makes the growth rate small for fixed value of
$\xi$, resonance is most efficient in the case of $g=0$.

%%%%%%%%%%%%%%%%%%%%%%%%%%%
\subsection{summary}    %
%%%%%%%%%%%%%%%%%%%%%%%%%%%

We shall summarize the properties of the resonance
and the suppression effect.
On the $\xi$-$g$ diagram (Fig.~11), we show which suppression
factor is significant in various coupling regimes.
We can easily find that the back reaction effect is dominant
in the case of $g~\mbox{\raisebox{-1.ex}{$\stackrel
     {\textstyle>}{\textstyle\sim}$}}~3 \times 10^{-4}$ and $g~\mbox{\raisebox{-1.ex}{$\stackrel
     {\textstyle>}{\textstyle\sim}$}}~5 \times
10^{-6} |\xi|$ (the lined region in Fig.~11). However, in the
case of
$g~\mbox{\raisebox{-1.ex}{$\stackrel
     {\textstyle<}{\textstyle \sim}$}}~3
\times 10^{-4}$ and $|\xi| \gg 1$, the $\xi\eta$ suppression
effect becomes more important. In fact, $\xi\eta$ suppression
appears either for $\xi~\mbox{\raisebox{-1.ex}{$\stackrel
     {\textstyle>}{\textstyle\sim}$}}~100$ or
$\xi/g^2~\mbox{\raisebox{-1.ex}{$\stackrel
     {\textstyle<}{\textstyle \sim}$}}~-0.1
(M_{\rm PL}/m)^2$ (the shaded region in Fig.~11). In the
parameter range like
 $-(M_{\rm PL}/m)^2/12\pi  <\xi/g^2 <- (M_{\rm
PL}/m)^2/16\pi$ (new-broad resonance with $\xi<0$ case),
however, although
$|\xi|$ is of order 100, the  effective $\chi$-particle
production will not be expected.
In the region where the back reaction or suppression effect
becomes significant, we may expect a large amount of particle
production, although it will be reduced for extremely large
$|\xi|$.

In Fig.~12, we show $\langle \bar{\chi}^2 \rangle_{f}$ in terms
of $\xi$ and $g$  in 3-dimensional plane.
{}From this figure, we find three plateaus; one corresponds to
the back reaction region with large $g$, and the other two
correspond to  $\xi\eta$ suppression regions shown in Fig.~11.
$\langle\chi^2\rangle_f$ takes maximal value of 
$\langle \bar{\chi}^2\rangle_{max} \approx 3.2
\times 10^8$ at $g~\mbox{\raisebox{-1.ex}{$\stackrel
{\textstyle<}{\textstyle \sim}$}}~1 \times 10^{-5},
\xi \approx -4$.
In other regions, we find that the numerical result agrees
well with our analysis.

In this section, we have studied the combined resonance of $g$
and $\xi$. For $g~\mbox{\raisebox{-1.ex}{$\stackrel
 {\textstyle>}{\textstyle\sim}$}}~ 3 \times 10^{-4}$, when the back reaction effect is important, the final values of $\langle \chi^2 \rangle$ does not become larger than those of $g$-resonance
significantly even if $\xi$-coupling is taken into account.
This is due to both of the $\xi\eta$ suppression  and the
back reaction. In the case that $|\xi|$ is large  as
$|\xi|~\mbox{\raisebox{-1.ex}{$\stackrel{\textstyle>}{\textstyle\sim}$}}~100$, the $\xi\eta$ suppression effect becomes more important than the back reaction.
For $g~\mbox{\raisebox{-1.ex}{$\stackrel{\textstyle<}{\textstyle \sim}$}}~ 3 \times 10^{-4}$, $g$-resonance is sometimes assisted by $\xi$-coupling.
For example, for $100~\mbox{\raisebox{-1.ex}{$\stackrel
     {\textstyle<}{\textstyle \sim}$}}~\xi~\mbox{\raisebox{-1.ex}{$\stackrel
     {\textstyle<}{\textstyle \sim}$}}~200$ and $-\infty
<\xi/g^2<-(M_{\rm PL}/m)^2/16\pi$, $\xi$-coupling  assists
$g$-resonance. In particular, for $g~\mbox{\raisebox{-1.ex}{$\stackrel
     {\textstyle<}{\textstyle \sim}$}}~1 \times 10^{-5}$,
the resonance structure is almost the same as the case of
$g=0, \xi
\ne 0$, which means that it is essentially the $\xi$-resonance.

%%%%%%%%%%%%%%%%%%%%%%%%%%%%%%%%%%%%%%%%%%%%%%%%%%
%%%%%%%%%%
%                                                          %
\section{Conclusions and Discussions}    %
%                                                          %
%%%%%%%%%%%%%%%%%%%%%%%%%%%%%%%%%%%%%%%%%%%%%%%%%%
%%%%%%%%%%

In this paper we have examined the properties of resonance
with non-minimally coupled scalar field $\chi$  in preheating
phase.
We have found that effective resonance is possible only  by a
non-minimal coupling $\xi R \chi^2$ in a certain
range of parameter
$\xi$. In the case of $g=0, \xi>0$, the relation of $A_k$ and
$q$ in the  Mathieu chart is $A_k =2q/3+ \bar{k}^2/{a^2}$ and
the resonance band becomes broader in comparison with
$g$-resonance ($A_k=2q+
\bar{k}^2/{a^2}$).
This is due to the existence of $\xi\kappa^2 \dot{\phi}^2$ term
and the structure of resonance is different from that of
$g$-resonance. Without $\xi\kappa^2 \dot{\phi}^2$ term, $\xi$
must be larger  than $10^4$ for the  effective resonance.
However, as we have shown here,  the  effective resonance is
possible for
$\xi \sim O(100)$. For example in the case of
$g=0, \xi=100$, we find
$\sqrt{\langle \chi^2\rangle}_f \approx 10^{16}$ GeV,
which is comparable to the case of $g \approx 3 \times 10^{-4},
\xi=0$. The unique feature of $\xi$-resonance is the existence
of
$\xi\eta$  suppression effect.
As $\xi$ increases up to about 100, $\langle \chi^2\rangle_f$
also increases because  $\chi$-field stays
in the broad resonance bands longer.
However, when $\xi$ exceeds about 100,  $\xi\eta$ suppression
effect by the
 production of $\chi$-particles is significant and $\langle
\chi^2
\rangle_f$ does not increase for the case of 
$\xi~\mbox{\raisebox{-1.ex}{$\stackrel
     {\textstyle>}{\textstyle\sim}$}}~100 $.
Rather, beyond  $\xi~\sim~200$, $\langle \chi^2\rangle_f$
decreases by $\xi\eta$ suppression effect as
$\langle \chi^2 \rangle_f \propto \xi^{-3/2}$.
In the case of $g=0, \xi>0$, we find that the maximal value of
$\langle
\chi^2 \rangle_f$ is  $\sqrt{\langle \chi^2 \rangle}_{max}
\approx
10^{16}$ GeV at $\xi=100 \sim 200$.

As for the case of $g=0, \xi<0$, the relation between $A_k$ and
$q$  becomes
$A_k=-2q/3+\bar{k}^2/{a^2}$ and the resonance band is further
broader than the case of $g=0, \xi>0$.
The important difference from other cases is that $A_k$-$q$
curve will pass through the zeroth instability band
below the curve of
$A_k=-q^2/2$. As a result, even if $q$ decreases under unity by
expansion of the  Universe, the modes close to $k=0$
always stay in the resonance band. In this case, we  find the
termination in  the growth of
$\langle \chi^2\rangle$ at which the growing rate $\mu$ of
$\langle
 Y^2\rangle$ balances the expansion rate of the Universe.
In $-1~\mbox{\raisebox{-1.ex}{$\stackrel
     {\textstyle<}{\textstyle \sim}$}}~\xi~\mbox{\raisebox{-1.ex}{$\stackrel{\textstyle<}{\textstyle \sim}$}}~0$ case,
the increase of $\langle\chi^2\rangle$ is not expected because
the Hubble expansion rate surpasses the growing rate.
In $-3~\mbox{\raisebox{-1.ex}{$\stackrel{\textstyle<}{\textstyle \sim}$}}~\xi~\mbox{\raisebox{-1.ex}{$\stackrel{\textstyle<}{\textstyle \sim}$}}~-1$ case, although $\langle\chi^2\rangle$ increases
as the passage of time, it takes much time to reach its
maximum because $\mu$ is very small and resonance terminates
before $\xi\eta$ suppression effect becomes significant.
For $\xi~\mbox{\raisebox{-1.ex}{$\stackrel{\textstyle<}{\textstyle \sim}$}}~-3$,  $\langle \chi^2\rangle_f$ takes rather large values
and $\langle \chi^2\rangle_f$ takes maximal 
value $\sqrt{\langle \chi^2\rangle}_{max} \sim 2 \times 10^{17}$ GeV 
when $\xi \approx -4$.
 As $\xi$ decreases from $\xi~\mbox{\raisebox{-1.ex}{$\stackrel{\textstyle<}{\textstyle \sim}$}}~-4$, although the growth rate $\mu$ increases, the final fluctuation of $\chi$-particle decreases.
This is because $\xi\eta$ suppression effect
due to $\chi$-particle production plays a crucial role to terminate
the resonance.
As a result, $\langle \chi^2\rangle_f$ decreases as $|\xi|^{-3/2}$
for $|\xi| \gg 1$.
We should also note  that the value of $\sqrt{\langle
\chi^2 \rangle}_{max}$ in the case of $g=0, \xi<0$ is greater than
that in the case of $g=0, \xi>0$.
This is because the resonance bands for $\xi <0$ are broader
than those for $\xi>0$.

We have also studied the combined resonance of interactions $
g^2\phi^2\chi^2$ and $\xi R \chi^2$.
The structure of resonance is quite different depending on two
parameters of $ g$ and $\xi$.
What we had been interested in is whether $\xi$-coupling
assists
$g$-resonance in any cases.  However, we find that
this is not the
case. For $g~\mbox{\raisebox{-1.ex}{$\stackrel
     {\textstyle>}{\textstyle\sim}$}}~3 \times 10^{-4}$,
and back reaction effect
on the inflaton field and metric is important, $g$-resonance is
not assisted  by $\xi$-coupling because of $\xi\eta$
suppression effect as well as back  reaction effect at the
final stage of preheating. In the parameter range where
$\chi$-particles are significantly produced  only by
$g$-resonance, the maximal value of
$\langle \chi^2\rangle_f$ does not increase even if we include
the
$\xi$-coupling.
On the other hand, in the case of $g~\mbox{\raisebox{-1.ex}{$\stackrel
     {\textstyle<}{\textstyle \sim}$}}~3 \times 10^{-4}$,
$\xi$-coupling may assist
 $g$-resonance in the parameter ranges of
$100~\mbox{\raisebox{-1.ex}{$\stackrel
     {\textstyle<}{\textstyle \sim}$}}~\xi~\mbox{\raisebox{-1.ex}{$\stackrel
     {\textstyle<}{\textstyle \sim}$}}~200$ and $-\infty <\xi/g^2<-
\left(  M_{\rm PL}/m \right)^2/16\pi $.
In particular,  for $g~\mbox{\raisebox{-1.ex}{$\stackrel
     {\textstyle<}{\textstyle \sim}$}}~1 \times 10^{-5}$, the structure
of resonance is almost the same as the case with $g=0, \xi \ne
0$, and it is essentially the $\xi$-resonance.
We find that the maximal value of $\sqrt{\langle
\chi^2\rangle}_f$ is about $2 \times 10^{17}$ GeV for $g~\mbox{\raisebox{-1.ex}{$\stackrel{\textstyle<}{\textstyle \sim}$}}~1 \times 10^{-5}, \xi \approx
-4$, which is larger than the minimally coupled case with
$g \approx 1 \times 10^{-3}$.

There are several things we did not investigate in this paper.
One of them is the rescattering effect.
As $\chi$-particles are produced significantly, the
fluctuations of the inflaton field are also generated and would
affect the production of $\chi$-particles.
Although it is expected that
the structure of resonance does not change so much at the
first stage of preheating, the rescattering effect will modify  
the final production of $\chi$-particles
because it becomes important
at the final stage of preheating when $\chi$-particles are
significantly produced.
As $\phi$-particles are created, 
spatially inhomogeneity of the $\phi$-field would prevent from the
resonant  production of $\chi$-particles, resulting
$\eta=\xi\kappa^2\langle\chi^2\rangle$ to be reduced. Then we
may find either insufficient production of $\chi$-particles or the
same amount of $\chi$-particles with a delay of time to reach the
limiting value 
due to the $\xi\eta$ suppression.
As for $\phi$-particle creation through the present coupling,
we did not investigate details at present.
For complete study of preheating, however, we should consider
the growth of $\phi$-particles due to rescattering effect
quantitatively beyond mean field approximation.

We did not consider the case when the coupling
$g^2$ between $\phi$ and $\chi$ is negative, although it was
pointed out in \cite{BPR} that the production of $\chi$-particles
increases significantly  compared with the positive coupling
case. This is similar to the geometric reheating with negative
$\xi$ in the sense that
$A_k$ can take a negative value in the Mathieu chart.
It may be interesting to study the combined resonance of
negative $g^2$ and $\xi$.

In this paper, we have also not studied the
metric perturbation in preheating phase.
However, several authors pointed out that metric perturbation
is  influenced by the  parametric resonance
\cite{mperturbation1,mperturbation2,mperturbation3,mperturbation4}.
It was recognized the Bardeen parameter is well conserved
quantity in reheating phase except the short period when
$\dot{\phi}$ is close to zero\cite{mperturbation1,mperturbation4}.
On the other hand, Bassett et al. \cite{mperturbation3}
recently found that the rapid growth of metric perturbation by
negative  coupling instability is expected and this
stimulates the  growth of scalar field.
It is worth investigating whether the growth of
metric perturbation enhances the fluctuation of $\chi$ field
non-minimally  coupled to spacetime curvature $R$ with $\xi <0$.

We have studied  a parametric resonance  by 
$\xi$-coupling ($\xi R\chi^2$) as well as
$g$-interaction ($g^ 2\phi^2\chi^2$).
Although we expect that any scalar field will couple to
spacetime curvature $R$ through quantum effects, the value
of $\xi$
considered here may be too large.
However, in other theories of gravity such as
the Brans-Dicke theory\cite{BD}, the induced
gravity\cite{induced}, and the higher-curvature
theories\cite{higher_curvature}, we may have different types of
coupling to the spacetime curvature, which might give a natural
mechanism for an effective resonance.
These issues are under investigation.

%%%%%%%%%%%%%%%%%%%%%%%%%%%%%%%%%%%%%%%%%%%%%%%%%%
\section*{ACKOWLEDGEMENTS}
We would like to thank Jun-ichirou Koga and Hiroki Yajima
for useful discussions.
T. T. is thankful for financial support from the JSPS. This
work was supported partially by a Grant-in-Aid for  Scientific
Research Fund of the Ministry of Education, Science and Culture
(No. 09410217 and Specially Promoted Research No. 08102010), by a
JSPS Grant-in-Aid
(No. 094162), and by the Waseda University Grant  for Special
Research Projects.

\newpage
%%%%%%%%%%%%%%%%%%%%%%%%%%%%%%%%%%%%%%%
%%%%%%%%%%%%%%%%%%%%%%%%%%%%%%%%%%%%%%%

\newpage

\begin{table}
\caption{The final values $\langle \bar{\chi}^2 \rangle_{f}$
obtained by the analytical estimation and by the numerical
calculation in the case of $g=0, \xi>0$. The analytical
estimation  gives a good approximation to the numerical
results.
$\langle \bar{\chi}^2 \rangle_{numerical}$ takes
the maximal value  at $\xi \sim 100$. We also show the time
$\bar{t}_f$ when the resonance ceases and $\xi\eta$ which
indicates a suppression effect. For large $\xi$, the suppression
effect ($q
\sim (1+6\xi\eta)^{-2}$) becomes  crucial.}

\vskip .3cm
\noindent
\begin{tabular}{crc|cc|ccc|lc}
 ~& $\xi$ & ~& $\bar{t}_{f} $ &~&
         $\langle \bar{\chi}^2 \rangle_{analytic}$
         & $\langle \bar{\chi}^2 \rangle_{numerical}$
   &~& $\xi \eta_{numerical}$ &~\\
        \hline
 ~&20 & ~& 1.85 & ~&--- & 8.610   &~ & $8.7 \times 10^{-8}$
&~\\
 ~& 50 & ~& 2.85 & ~&---  & $3.090 \times 10^3$  &~ & $1.9
\times 10^{-4}$  &~\\   ~&70 & ~& 2.85 & ~&---   & $3.750 \times
10^5$  &~ & $4.6 \times 10^{-2}$  &~\\   ~&100 & ~& 2.40 &~&
$4.138 \times 10^5$   & $9.550 \times 10^5$  &~ &
 $2.4 \times 10^{-1}$ &~\\
 ~&500 & ~& 1.31 &~& $1.513 \times 10^5$   & $2.109 \times
10^5$ &~ & 1.3 &~\\
 ~&1000 & ~&1.31 &~& $5.626 \times10^4$  & $8.318 \times10^4$
&~ & 2.1  &~\\
 ~&10000 & ~&1.31 &~& $1.922 \times10^3$ & $2.630 \times10^3$
&~& 6.6  &~
\end{tabular}
\end{table}
%%%%%%%%%%%%%%%%
\vspace{1cm}
\begin{table}
\caption{The final value $\langle \bar{\chi}^2 \rangle_{f}$
obtained by
the analytical estimation and by the numerical
calculation in the case of $g=0, \xi<0$.
For $|\xi| \ge 4$, both values show the same tendency that
$\langle \bar{\chi}^2 \rangle_{f}$ gets large for decreasing
$|\xi|$. However, for $|\xi| \le 3$,
the analytic estimation by Eq.~(4.8) can not be applied
(see  text for the detail).
$\langle \bar{\chi}^2 \rangle_{numerical}$ takes
the maximal value at $\xi \approx -4$.
}

\vskip .3cm
\noindent
\begin{tabular}{crc|cc|ccc|lc}
 ~&      $\xi$ &~& $\bar{t}_{f} $ &~&
         $\langle \bar{\chi}^2 \rangle_{analytic}$
         & $\langle \bar{\chi}^2 \rangle_{numerical}$
         &~& $|\xi |\eta_{numerical}$ &~\\
        \hline
~&$-2$ & ~& $7.5 \times 10^5$ &~&  $1.049 \times 10^9$   & $3.184 \times 10^3$   &~& $3.2 \times 10^{-7}$  &~\\
~&$-4$ & ~&$3.3 \times 10^5$ &~&  $5.425 \times 10^8$   & $3.162  \times 10^8$   &~& $1.3 \times 10^{-1}$  &~\\
 ~&$-10$ & ~&$1.0 \times 10^3$ &~&  $1.758 \times 10^8$   & $1.122  \times 10^8$   &~& $2.8 \times 10^{-1}$  &~\\
 ~&$-20$ & ~& 28.28 &~& $6.903 \times10^7$  & $4.130  \times
10^7$  &~& $4.2 \times 10^{-1}$ &~\\
 ~&$-50$ & ~& 15.26 &~& $1.896 \times10^7$ & $1.099  \times
10^7$    &~& $6.9 \times 10^{-1}$ &~\\
 ~&$-100$ & ~& 7.66 &~& $6.994 \times10^6$ & $4.355 \times 10^6$
&~& 1.1 &~\\
 ~&$-1000$ & ~& 4.61 &~& $2.355 \times10^5$ & $1.606  \times
10^5$   &~& 4.0 &~\\
 ~&$-10000$ & ~& 4.18 &~& $7.591 \times10^3$ & $5.069  \times
10^3$   &~& 12.7 &~\\
\end{tabular}
\end{table}
%%%%%%%%%%%%%%%%
\vspace{1cm}
\begin{table}
\caption{The numerical result for the
broad resonance in the $g = 3 \times 10^{-4}$ case.
$\bar{t}_{b} $ and $\langle \bar{\chi}^2 \rangle_{b}$ are the
values when the broad resonance regime ends.
$\langle \bar{\chi}^2 \rangle_{b}$  gets larger for large
$\xi$ because the resonance band becomes very broad.
However,  $\langle \bar{\chi}^2 \rangle_f$
turns out to be smaller for large $\xi$ because of the
suppression  effect, although it does not depend on $\xi$ so
much.}

\vskip .3cm
\noindent
\begin{tabular}{crc|ccc|ccc}
        ~& $\xi$ & ~&
         $\bar{t}_{b} $ &
         $\langle \bar{\chi}^2 \rangle_{b}$ &~&
         $\bar{t}_{f} $ &
         $\langle \bar{\chi}^2 \rangle_{f}$
         &~\\
        \hline
~&$-20$ &~& 6.26 & $3.019$  &~& 14.36  & $2.951 \times 10^5$
&~\\ ~&$ 0$ &~& 6.26 & $5.495$ &~& 14.36 & $7.244 \times 10^5$
&~ \\ ~&$50$ &~& 6.26 & $3.162 \times10$ &~& 12.85  & $6.309
\times 10^5$  &~ \\ ~&$100$ &~& 6.26 & $2.344 \times10^{2}$ &
~&11.88  &
$4.570 \times 10^5$  &~ \\
\end{tabular}
\end{table}
%%%%%%%%%%%%%%%%
\vspace{1cm}
\begin{table}
\caption{The numerical result for the
new-broad resonance on the $A_k=2q/3$ line on the Mathieu chart.
 $g$ does not take large value for this parameter
range of $\xi$. The effective resonance is expected only for
$|\xi| >200$.
}

\vskip .3cm
\noindent
\begin{tabular}{crc|ccc}
        ~& $\xi$ & ~&
         $g$ &
         $\langle \bar{\chi}^2 \rangle_{f}$
         &~\\
        \hline
~&$-20$ &~& $2.745 \times10^{-5}$& ---   &~\\
~&$-50$ &~& $4.341 \times10^{-5}$& ---    &~\\
~&$-70$ &~& $5.137 \times10^{-5}$& ---    &~\\
~&$-100$ &~& $6.140 \times10^{-5}$& $5.012 \times10$   &~\\
~&$-200$ &~& $7.089 \times10^{-5}$& $1.013 \times 10^5$   &~\\
\end{tabular}
\end{table}
%%%%%%%%%%%%%%%
\vspace{1cm}
\begin{table}
\caption{The numerical result for the
wide resonance on the $A_k=0$ line on the
Mathieu chart. The resonance occurs mainly by the $\xi$-coupling.
The effective resonance is expected only for  $|\xi|>100$.
}

\vskip .3cm
\noindent
\begin{tabular}{crc|ccc}
~&       $\xi$ &~&
         $g$ &
         $\langle \bar{\chi}^2 \rangle_{f}$
         &~\\
        \hline
~&$-20$ &~&$ 2.241 \times10^{-5}$& ---   &~\\
~&$-50$ &~& $3.545 \times10^{-5}$& $3.981$    \\
~&$-70$ &~& $4.194 \times10^{-5}$& $1.016 \times 10^4$  &~ \\
~&$-100$ &~& $5.013 \times10^{-5}$& $3.206 \times 10^5$ &~\\
~&$-200$ &~& $7.089 \times10^{-5}$& $1.047 \times10^6$  &~\\
\end{tabular}
\end{table}

\newpage
%%%%%%%%%%%%%%%%%%%%
%   figures
%%%%%%%%%%%%%%%%%%%%
\begin{flushleft}
{ Figure Captions}
\end{flushleft}
\noindent
%%%%%%%%
\parbox[t]{2cm}{FIG. 1:\\~}\ \
\parbox[t]{12cm}
{The schematic diagram of the Mathieu chart and the typical
paths for three types of resonance.  The lined regions
denote  the  instability bands (zeroth, first, second,$\cdots$).
 The line of $A_k=2q$ is the
typical line of the ordinary
$g$-resonance, while the lines of  $A_k=2q/3$ and $A_k=-2q/3$ show
the lowest limits of the resonances by positive  and
negative
$\xi$-couplings, respectively. We  find that the width of the
instability bands becomes wider for large $q$. The Floquet index
in the lower instability band gets larger for fixed $q$.

\label{solution}  }\\[1em]
\noindent
%%%%%%%%
%%%%%%%%
\parbox[t]{2cm}{FIG. 2:\\~}\ \
\parbox[t]{12cm}
{The evolution of $\langle \bar{\chi}^2 \rangle$ as a function of
$\bar{t}$ in the case of $g=0, \xi>0$ ((a)~$\xi=50$,
(b)~$\xi=100$ and (c)~$\xi=1000$). We   find that the
parametric resonance occurs by the positive
$\xi$-coupling. $\langle \bar{\chi}^2 \rangle$ increases
exponentially and reaches to its final value
$\langle \bar{\chi}^2 \rangle_f$. After then, it decreases by the
adiabatic expansion.
$\langle \bar{\chi}^2 \rangle_f$ takes the maximal value for
$\xi \sim 100$. For larger $\xi$ ($\xi~\mbox{\raisebox{-1.ex}{$\stackrel
     {\textstyle>}{\textstyle\sim}$}}~100$), although
the growth rate becomes large, the final value
$\langle \bar{\chi}^2 \rangle_f$ is suppressed by the $\xi\eta$
term.
}\\[1em]
\noindent
%%%%%%%%
%%%%%%%%
\parbox[t]{2cm}{FIG. 3:\\~}\ \
\parbox[t]{12cm}
{The final value of $\langle \bar{\chi}^2 \rangle$ as a function
of
$\xi$ in the case of  $g=0, \xi>0$.
For $0~\mbox{\raisebox{-1.ex}{$\stackrel
     {\textstyle<}{\textstyle \sim}$}}~\xi~\mbox{\raisebox{-1.ex}{$\stackrel
     {\textstyle<}{\textstyle \sim}$}}~100$, $\langle \bar{\chi}^2 \rangle_f$
increases  as $\xi$  increases because the initial value of $q$
gets
 large. For $100~\mbox{\raisebox{-1.ex}{$\stackrel
     {\textstyle<}{\textstyle \sim}$}}~\xi~\mbox{\raisebox{-1.ex}{$\stackrel
     {\textstyle<}{\textstyle \sim}$}}~200$, $\langle \bar{\chi}^2
\rangle_f$
 takes its maximal value ($\sim 10^6$), and
for $\xi~\mbox{\raisebox{-1.ex}{$\stackrel
     {\textstyle>}{\textstyle\sim}$}}~200$, $\langle \bar{\chi}^2 \rangle_f$ decreases
by the suppression factor as $\sim \xi^{-3/2}$.}\\[1em]
\noindent
%%%%%%%%
%%%%%%%%
\parbox[t]{2cm}{FIG. 4:\\~}\ \
\parbox[t]{12cm}
{The evolution of $\langle \bar{\chi}^2 \rangle$ as a function of
$\bar{t}$ in the case of $g=0, \xi=200, m_{\chi} =m$.
The significant production of massive $\chi$-particles of order
$m \sim 10^{13}$ GeV is expected by the   positive
$\xi$-resonance.
}\\[1em]
\noindent
%%%%%%%%
%%%%%%%%
\parbox[t]{2cm}{FIG. 5:\\~}\ \
\parbox[t]{12cm}
{The evolution of $\langle \bar{\chi}^2 \rangle$ as a function of
$\bar{t}$ in the case of $g=0, \xi<0$ (
(a)~$\xi=-20$ , (b)~$\xi=-50$ and (c)~$\xi=-100$).
$\langle \bar{\chi}^2 \rangle$ increases
exponentially in the first stage by the parametric
resonance and reaches to its final value
$\langle \bar{\chi}^2 \rangle_f$. After then, it approaches
constant value because the creation rate of $\chi$-particle and
the expansion rate of the Universe balance.
}\\[1em]
\noindent
%%%%%%%%
%%%%%%%%
\parbox[t]{2cm}{FIG. 6:\\~}\ \
\parbox[t]{12cm}
{The evolution of $\langle \bar{\chi}^2 \rangle$ as a function of
$\bar{t}$ in the case of $g=0, \xi=-200, m_{\chi} =m$.
The final value of $\langle \bar{\chi}^2 \rangle_f \approx
10^{6.5}$ is larger
 than the case of  $g=0, \xi=200, m_{\chi} =m$. }\\[1em]
\noindent
%%%%%%%%
%%%%%%%%
\parbox[t]{2cm}{FIG. 7:\\~}\ \
\parbox[t]{12cm}
{The final value of $\langle \bar{\chi}^2 \rangle$ as a function
of $\bar{m}_{\chi}=m_{\chi}/m$ in the case of $\xi=-1000$.
$\langle \bar{\chi}^2 \rangle_f$ decreases as $\bar{m}_{\chi}$
increases, and parametric resonance can not be expected
for $\bar{m}_{\chi}~\mbox{\raisebox{-1.ex}{$\stackrel
     {\textstyle>}{\textstyle\sim}$}}~100$ (namely, for
$m_{\chi}~\mbox{\raisebox{-1.ex}{$\stackrel
     {\textstyle>}{\textstyle\sim}$}}~10^{15}$~GeV). }\\[1em]
\noindent
%%%%%%%%
%%%%%%%%
\parbox[t]{2cm}{FIG. 8:\\~}\ \
\parbox[t]{12cm}
{The relation between $A_k$ and $q$ in the case of the combined
resonance of $g$ and $\xi$.
(a) As $\xi/g^2$ decreases from $\infty$ to
$-(M_{\rm PL}/m)^2/24\pi$, the relation
changes from $A_k=2q/3+\bar{k}^2/a^2$ to $q=0$  as the arrow in
the figure.  We call the resonance in each parameter region as
follows.
$0<\xi/g^2<\infty$ : new-broad resonance, and
$-(M_{\rm PL}/m)^2/24\pi<\xi/g^2<0$ : ordinary-broad resonance.
(b) As $\xi/g^2$ decreases from
$-(M_{\rm PL}/m)^2/24\pi$
to $-\infty$, the relation changes from
$q=0$ to $A_k=-2/3 q+\bar{k}^2/a^2$ as the arrow in the figure.
We call the resonance in each parameter region as follows.
$-(M_{\rm PL}/m)^2/16\pi<\xi/g^2<-(M_{\rm PL}/m)^2/24\pi$ :
ordinary-broad resonance,
$-(M_{\rm PL}/m)^2/12\pi<\xi/g^2<-(M_{\rm PL}/m)^2/16\pi$ :
new-broad resonance, and
$-\infty<\xi/g^2<-(M_{\rm PL}/m)^2/12\pi$ : wide
resonance. }\\[1em]
\noindent
%%%%%%%%
%%%%%%%%
\parbox[t]{2cm}{FIG. 9:\\~}\ \
\parbox[t]{12cm}
{The evolution of $\langle \bar{\chi}^2 \rangle$ as a function
of
 $\bar{t}$ in the case of $g=3.0\times10^{-4}$( (a)~$\xi=0$,
(b)~$\xi=50$  and (c)~$\xi=100$). We  find that the
resonance is divided into two stages; one of which is the first
broad resonance stage and the other is the narrow one. The growth
rate in the broad resonance stage becomes larger for larger
$\xi$. However the final value $\langle \bar{\chi}^2 \rangle_f$
is  suppressed by the $\xi\eta$ term.
}\\[1em]
\noindent
%%%%%%%%
%%%%%%%%
\parbox[t]{2cm}{FIG. 10:\\~}\ \
\parbox[t]{12cm}
{The evolution of $\langle \bar{\chi}^2 \rangle$ as a function of
$\bar{t}$ in the case of  $g=1.0\times10^{-3}$( (a)~$\xi=0$,
(b)~$\xi=50$  and (c)~$\xi=100$). The broad
and  narrow resonance stages  are not  distinguished.
The growth rates are almost the same for each
$\xi$, because the resonance occurs mainly  by $g$-coupling.
When $\langle \bar{\chi}^2 \rangle$ reaches its maximal value,
the back reaction effect terminates the resonance.
}\\[1em]
\noindent
%%%%%%%%
%%%%%%%%
\parbox[t]{2cm}{FIG. 11:\\~}\ \
\parbox[t]{12cm}
{The structure of resonance, the back reaction  and
$\xi\eta$  suppression effects in terms of $\xi$ and $g$.
The regions [A], [B], [C], [D] denote new-broad resonance
($\xi>0$), ordinary-broad resonance ($- ( M_{\rm PL}/m)^2/16\pi
<\xi/g^2 <0$), new-broad resonance ($-(M_{\rm PL}/m)^2/12\pi
 <\xi/g^2 <- (M_{\rm PL}/m)^2/16\pi$), wide
resonance ($-\infty <\xi/g^2<-( M_{\rm PL}/m)^2/12\pi$)
respectively.
With this diagram, we easily understand the resonance structure.
The lined regions ($g~\mbox{\raisebox{-1.ex}{$\stackrel
     {\textstyle>}{\textstyle\sim}$}}~3  \times 10^{-4}$ and $g~\mbox{\raisebox{-1.ex}{$\stackrel
     {\textstyle>}{\textstyle\sim}$}}~5
\times 10^{-6}  |\xi|$) denote those where the back reaction
effect is significant.
The shaded regions ($g~\mbox{\raisebox{-1.ex}{$\stackrel
     {\textstyle<}{\textstyle \sim}$}}~3 \times 10^{-4}$ and either
$\xi~\mbox{\raisebox{-1.ex}{$\stackrel
     {\textstyle>}{\textstyle\sim}$}}~100$
or $\xi/g^2~\mbox{\raisebox{-1.ex}{$\stackrel
     {\textstyle<}{\textstyle \sim}$}}~-0.1 (M_{\rm PL}/m)^2$)
denote those where $\xi\eta$ suppression effect becomes
important.
}\\[1em]
\noindent
%%%%%%%%
%%%%%%%%
\parbox[t]{2cm}{FIG. 12:\\~}\ \
\parbox[t]{12cm}
{$\langle \bar{\chi}^2 \rangle_{f}$
in terms of  $\xi$ and $g$. We find three plateaus, one of
which $\langle \bar{\chi}^2 \rangle_f$ takes maximal value
$\langle \bar{\chi}^2 \rangle_{max} \approx 3.2 \times 10^8$
for $g~\mbox{\raisebox{-1.ex}{$\stackrel{\textstyle<}{\textstyle \sim}$}}~1 \times 10^{-5}, \xi \approx -4$.
}\\[1em]
\noindent
%%%%%%%%


\begin{thebibliography}{99}
\bibitem{comment1} In the self-coupling model such as
$\lambda\phi^4$ interaction, a non-minimal coupling may not play
any important role\cite{Kaiser}.
\bibitem{comment2}
In the original frame (2.1), writing down the basic equations
and adopting the mean field approximation, we find its gravity part
as $(1-\xi\kappa^2\langle\chi^2\rangle) G_{\mu\nu}= \cdots$.
To write down this basic equation in Einstein frame, we have to adopt
the conformal factor  $\Omega^2=1-\eta$.
Although the basic equations in the original frame 
with the mean field approximation can be obtained
in above lines, those are very complicated.
Instead, we start with the basic equations in the Einstein frame
with the same conformal factor $\Omega^2=1-\eta$.
Although both basic equations
are not exactly the same because of how to treat 
derivative terms of the mean field approximation of 
the $\chi$ field, these terms
are not so important in our analysis.  
Then we expect that our results may still be reliable.
In order to perform exact analysis, we have to solve
the basic equations in the original frame by using lattice
simulation.
\end{thebibliography}
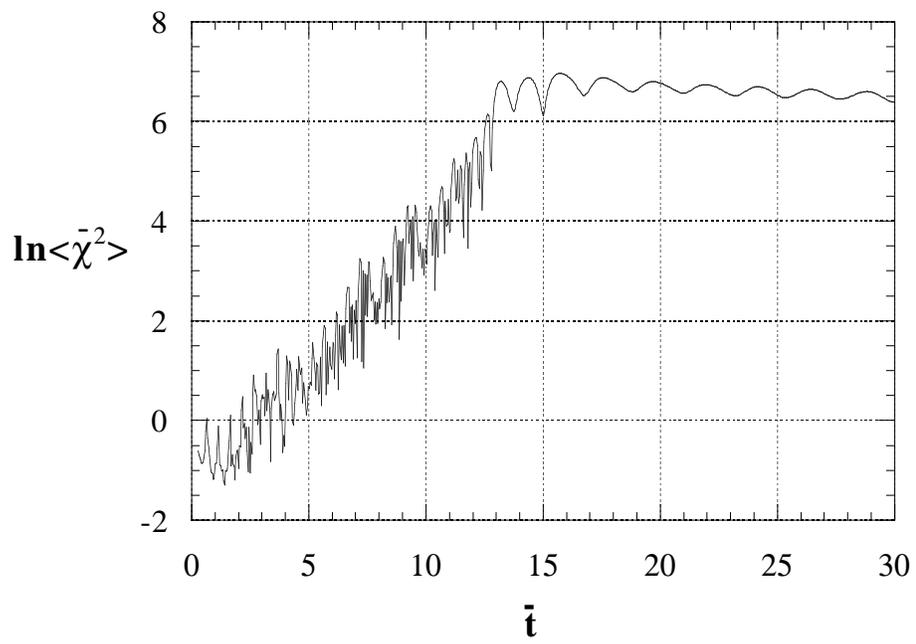
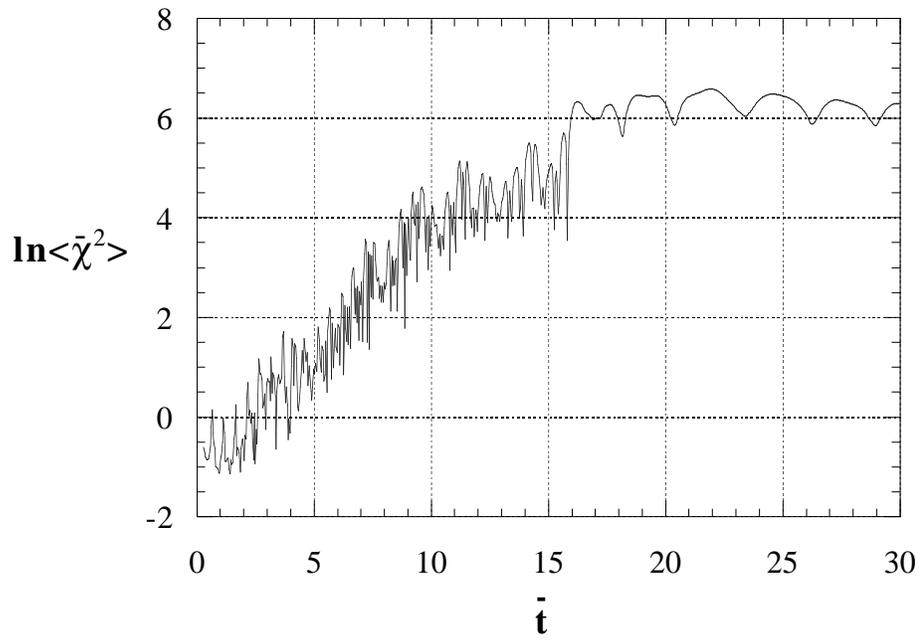
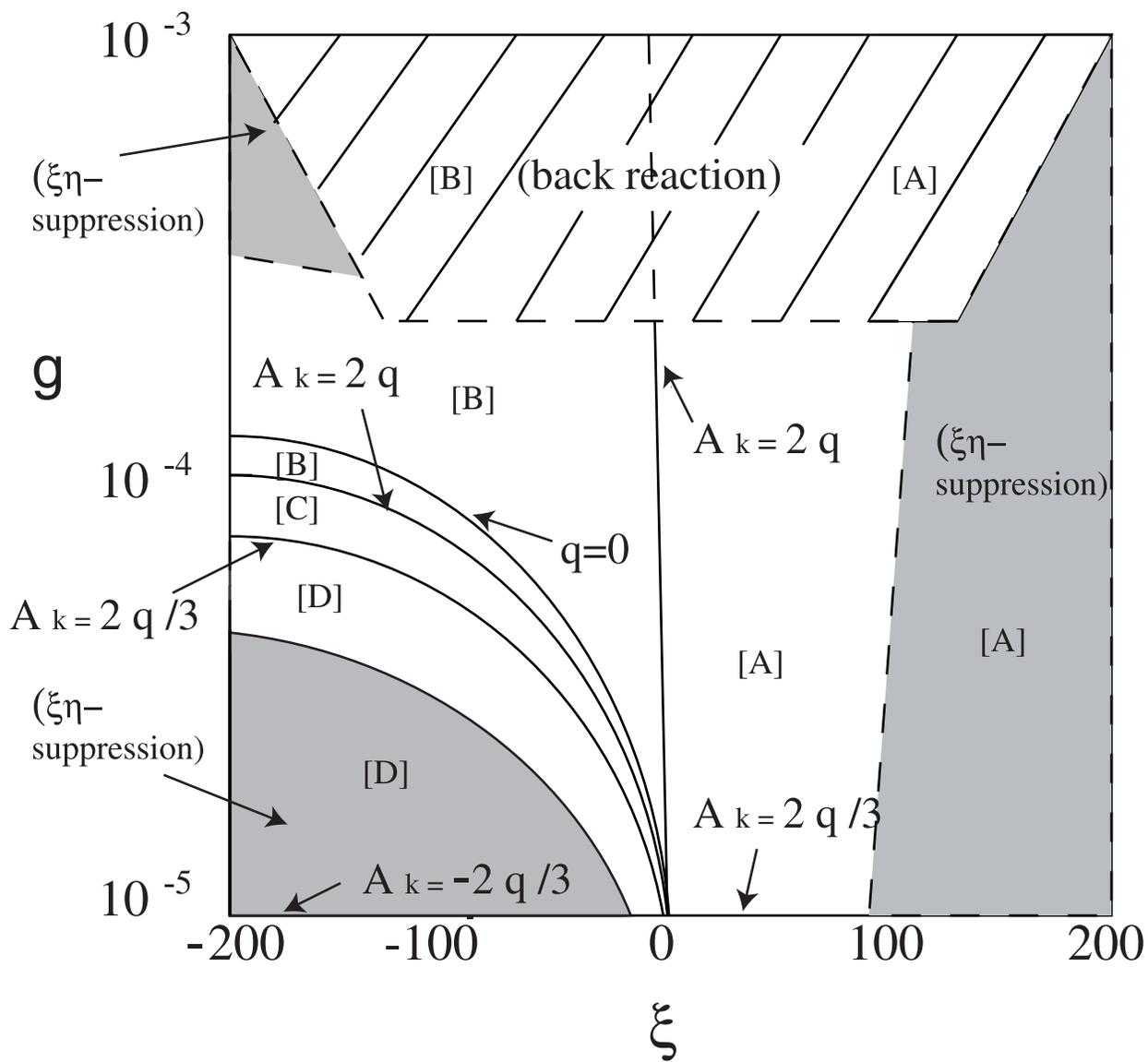
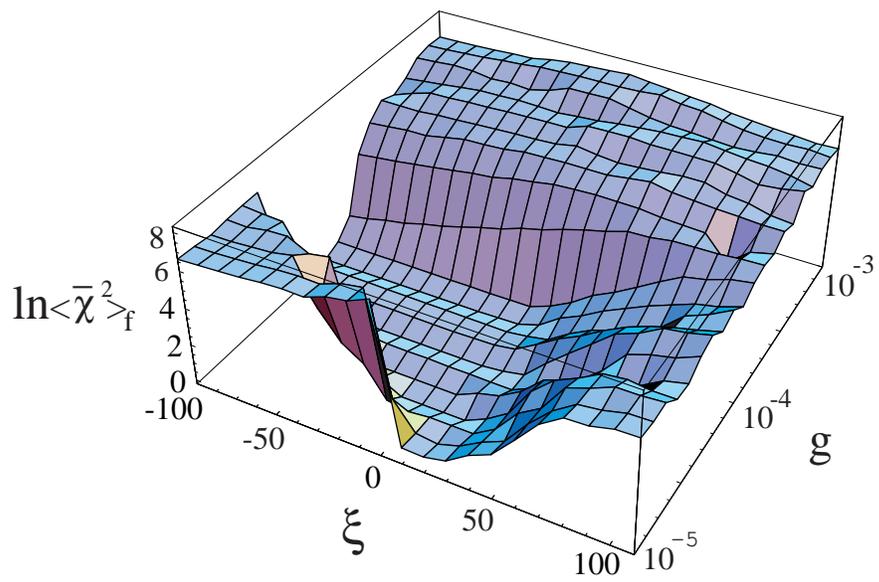
\end{document}